\documentclass[usenatbib,useAMS,usedcolumn]{mn2e}
\usepackage{longtable}
\usepackage{psfig}
\usepackage{epsfig}
\usepackage{graphicx}
\usepackage{times}
\def\arcm{\hbox{$^\prime$}}
\def\etal{{\rm et al.}\thinspace}
\def\eg{{\it e.g.\ }}
\def\etc{{\it etc.\ }}
\def\ie{{\it i.e.\ }}

\def\spose#1{\hbox to 0pt{#1\hss}}
\def\gtsim{$\mathrel{\spose{\lower 3pt\hbox{$\sim$}}
        \raise 2.0pt\hbox{$>$}}$\thinspace}
\def\ltsim{$\mathrel{\spose{\lower 3pt\hbox{$\sim$}}
        \raise 2.0pt\hbox{$<$}}$\thinspace}
\def\simpropto{$\mathrel{\spose{\lower 3pt\hbox{$\sim$}}
        \raise 2.0pt\hbox{$\propto$}}$\thinspace}
\newcommand{\rosat}{\emph{ROSAT}}
\newcommand{\chandra}{\emph{Chandra}}
\newcommand{\xmm}{\emph{XMM-Newton}}
\newcommand{\asca}{\emph{ASCA}}
\newcommand{\einstein}{\emph{Einstein}}
\newcommand{\arcs}{\mbox{\arcm\hskip -0.1em\arcm}}
\newcommand{\Lx}{\ensuremath{L_{\mathrm{X}}}}

\newcommand{\Zsol}{\ensuremath{Z_{\odot}}}
\newcommand{\Lsol}{\ensuremath{L_{\odot}}}
\newcommand{\Msol}{\ensuremath{M_{\odot}}}
\newcommand{\LB}{\ensuremath{L_{\mathrm{B}}}}
\newcommand{\LBsol}{\ensuremath{L_{B\odot}}}

\newcommand{\LxLb}{\ensuremath{\mbox{\Lx/\LB}}}

\newcommand{\Bfit}{\ensuremath{\beta_{fit}}}

\newcommand{\NH}{\ensuremath{N_{\mathrm{H}}}}

\newcommand{\s}{\ensuremath{\mbox{~s}}}
\newcommand{\ps}{\ensuremath{\s^{-1}}}
\newcommand{\km}{\ensuremath{\mbox{~km}}}
\newcommand{\Mpc}{\ensuremath{\mbox{~Mpc}}}
\newcommand{\pMpc}{\ensuremath{\Mpc^{-1}}}
\newcommand{\kmpspMpc}{\ensuremath{\km \ps \pMpc\,}}
\newcommand{\erg}{\ensuremath{\mbox{~erg}}}
\newcommand{\ergps}{\ensuremath{\erg \ps}}
\newcommand{\cm}{\ensuremath{\mbox{~cm}}}
\newcommand{\pcmsq}{\ensuremath{\cm^{-2}}}
\newcommand{\ergpcmsqps}{\ensuremath{\erg \pcmsq \ps}}
\newcommand{\kmps}{\ensuremath{\km \ps}}
\newcommand{\sas}{\textsc{sas}}
\newcommand{\ciao}{\textsc{ciao}}

\newcommand{\Ho}{\ensuremath{H_\mathrm{0}}}

\newcommand{\Dtf}{\ensuremath{D_{\mathrm{25}}}}

\newcommand{\Bspec}{\ensuremath{\beta_{spec}}}
\begin{document}

\title[
XMM-Newton and Chandra observations of three X-ray faint early-type galaxies
]
{
XMM-Newton and Chandra observations of three X-ray faint early-type galaxies
}
\author[
Ewan O'Sullivan et al. 
]
{
E. O'Sullivan$^1$\footnotemark, T.~J. Ponman$^2$\\
$^1$ Harvard Smithsonian Center for Astrophysics,
60 Garden Street, Cambridge, MA 02138, USA\\
$^2$ School of Physics and Astronomy, University of Birmingham, Edgbaston,
Birmingham, B15 2TT, UK\\
\\
}

\date{Accepted 2003 ?? Received 2003 ??; in original form 2003 ??}
\pagerange{\pageref{firstpage}--\pageref{lastpage}}
\def\LaTeX{L\kern-.36em\raise.3ex\hbox{a}\kern-.15em
    T\kern-.1667em\lower.7ex\hbox{E}\kern-.125emX}

\label{firstpage}

\maketitle

\begin{abstract}
  We present \xmm\ observations of three X-ray under-luminous elliptical
  galaxies, NGC~3585, NGC~4494 and NGC~5322. All three galaxies have
  relatively large optical luminosities (log \LB=10.35-10.67 \Lsol) but
  have X-ray luminosities consistent with emission from discrete sources
  only. In conjunction with a \chandra\ observation of NGC~3585, we analyse
  the \xmm\ data and show that the three galaxies are dominated by discrete
  source emission, but do possess some X-ray emitting gas. The gas is at
  relatively low temperatures, kT$\simeq$0.25-0.44 keV. All three galaxies
  show evidence of recent dynamical disturbance and formation through
  mergers, including kinematically distinct cores, young stellar ages, and
  embedded stellar disks. This leads us to conclude that the galaxies
  formed relatively recently and have yet to build up large X-ray halos.
  They are likely to be in a developmental phase where the X-ray gas has a
  very low density, making it undetectable outside the galaxy core.
  However, if the gas is a product of stellar mass loss, as seems most
  probable, we would expect to observe supersolar metal abundances. While
  abundance is not well constrained by the data, we find best fit
  abundances $<$0.1 \Zsol\ for single-temperature models, and it seems
  unlikely that we could underestimate the metallicity by such a large factor.
\end{abstract}


\footnotetext{Email: ejos@head-cfa.harvard.edu}

\section{Introduction}
\label{sec:intro}
Since the advent of the \einstein\ X-ray observatory, elliptical and
lenticular galaxies have been known to be luminous sources of X-ray
emission \citep{Formanjonestucker85}. In the most X-ray luminous examples,
typically found in the cores of groups and clusters, the majority of the
X-rays are produced through line emission and bremsstrahlung from hot
(10$^{6-7}$ K) gas located in extended halos surrounding and permeating the
galaxies \citep[\textit{e.g.}][]{Trinchierietal86}. The origins of these halos and the means by which they are
maintained are a subject of ongoing debate.  Possible sources of the gas
include stellar mass loss within the galaxy, infall and shock heating of
cool gas ejected during a merger \citep{HibvanG96}, and accretion of gas from a surrounding
cluster halo.

Another important source of X-rays in early-type galaxies is the X-ray
binary population. Prior to the launch of \chandra, little was known about
this component of the emission, there being no non-dwarf early-type 
galaxies in the
local group. \chandra, and to a lesser extent \xmm, have shown that stellar
sources can in fact dominate the emission from some early-type galaxies
\citep[\textit{e.g.}][]{SarazinIrBreg01,Blantonetal01,Baueretal01}. The
superb spatial resolution of \chandra\ permits the identification of large
numbers of individual sources in nearby galaxies, and their spectral
characterisation. The spectral properties of the unresolved source population 
can also be determined with some accuracy, thanks to the large
collecting areas of \chandra\ and \xmm.

In the case of relatively small, low mass galaxies it is to be expected
that X--ray binaries will be the dominant source of X--ray emission - the
shallow potential wells of these galaxies are less able to hold gas which
is thrown or blown outwards from within the galaxy, and less able to
accrete gas from outside. However, there are examples of optically luminous
ellipticals, which would normally be assumed to be fairly massive systems,
which are under-luminous in X-rays
\citep[\textit{e.g.}][]{Pellegrini94,Pellegrini99b,SarazinIrBreg01}. There
are several possible explanations for this. If these galaxies lack
significant dark matter halos, their stellar mass alone would be
insufficient to maintain large X-ray halos.  Alternatively, they might have
suffered from ram-pressure \citep{GunnGott72} or viscous \cite{Nulsen82}
stripping while passing through the dense core of a galaxy group or
cluster, removing most of their gas. In this case we would expect to see
nascent halos in the process of reconstruction. A third possibility is that
the galaxies have recently undergone a merger. Galaxy merging is observed
to trigger a burst of star formation, followed by the production of large
amounts of gas through supernovae and winds from massive stars
\citep[\textit{e.g.}][]{Readponman98}. Modelling of recent post-starburst
galaxies shows that they develop highly supersonic supernova driven winds
\citep{Ciottietal91,Pellegriniciotti98} whose low density leads to low
X-ray luminosity. This is in agreement with observations of post-merger
galaxies, which are known to be X-ray faint
\citep{MackieFabb96,Sansom00,OFP01age}.

We have used \xmm\ to observe three relatively nearby early-type galaxies,
NGC~3585, NGC~4494 and NGC~5322. The first two of these have also been
observed by \chandra\ and we combine these data with our own where
possible. These galaxies were chosen from the sample of \citet{OFP01cat} to
be optically luminous, with log \LB\ $\simeq$ 10.5 \LBsol, but with
exceptionally low \LxLb, close to the estimated typical luminosity from
stellar sources. We therefore expect these galaxies to have little or no
hot gas, and to be dominated by emission from point sources. Our intention
is to use these observations to shed light on the question of why their 
X-ray luminosity is so low. 

\begin{table}
\begin{center}
\begin{tabular}{lcccccc}
Galaxy & R.A. & Dec. & Dist. & 1\arcm= & \Dtf\ \\
 & (J2000.0) & (J2000.0) & (Mpc) & (kpc) & (kpc) \\
\hline
NGC 3585 & 11 13 17.1 & -26 45 18 & 21.28 & 4.67 & 13.14 \\
NGC 4494 & 12 31 24.0 & +25 46 30 & 16.07 & 6.19 & 13.83 \\
NGC 5322 & 13 49 15.2 & +60 11 26 & 27.80 & 8.09 & 24.36 \\
\end{tabular}
\end{center}
\caption{
\label{tab:basic}
Position and scale information for the galaxies in our sample. R.A. and
declination are taken from NED\protect\footnotemark[1], distances from the Fundamental
Plane study of \protect\citet{PrugnielSimien96}. \Dtf\ is the mean radius at
which the B-band optical intensity falls below 25 magnitudes per square
arcsecond, taken from LEDA\protect\footnotemark[2].
}
\end{table}
\footnotetext[1]{The NASA-IPAC Extragalactic Database,\\ http://nedwww.ipac.caltech.edu/}
\footnotetext[2]{The Lyon-Meudon Extragalactic Data Archive, http://leda.univ-lyon1.fr/}
%Footnote info at the end of section 2

Throughout the paper we assume \Ho=75\kmpspMpc\ and normalise optical
luminosities relative to the B-band optical luminosity of the sun,
\LBsol=5.2$\times$10$^{32}$ \ergps. In Section~\ref{sec:obs} we give
details of the observations and our basic analysis of the
data. Section~\ref{sec:res} details the results of spatial and spectral
fits to the galaxies, and in Section~\ref{sec:disc} we discuss the
implications of these results and draw conclusions.

%%%%%%%%%%%%%%%%%%%Table moved from below%%%%%%%%%%%%%
\begin{table*}
\begin{center}
\begin{tabular}{lcccccc}
Galaxy & \multicolumn{2}{c}{Exposure (ks)} & CCD Mode & Filter & Obs. date
& Counts in \Dtf \\
 & PN & MOS & & & & (BG subtracted) \\
\hline
\hline
\multicolumn{4}{l}{XMM-Newton}\\
\hline
NGC 3585 & 16.5 & 20.9 & PrimeFullWindow & Medium & 2001-12-27 & 3558\\
NGC 4494 & 24.0 & 29.5 & PrimeFullWindow & Medium & 2001-12-04 & 4457\\
NGC 5322 & 12.9 & 16.8 & PrimeFullWindow & Thin & 2001-12-24 & 2615\\
\hline
\multicolumn{4}{l}{Chandra}\\
\hline
NGC 3585 & \multicolumn{2}{c}{ACIS-S 32.7} & Faint & - & 2001-06-03 & 2394\\
\end{tabular}
\end{center}
\caption{
\label{tab:obs}
Details of the \xmm\ and \chandra\ observations of out target
galaxies. Exposures are effective, \ie\ the time remaining after removal of
flares and other bad time events. The last column shows the approximate
number of source counts (with the scaled background subtracted) within a
circle of radius \Dtf\ centred on the galaxy position.
}
\end{table*} 

\section{Observation and data reduction}
\label{sec:obs}
\subsection{\xmm\ observations}
The three galaxies were observed during XMM Cycle~1, using the EPIC
instruments. Details of the observations can be found in
Table~\ref{tab:obs}.  A detailed summary of the \xmm\ mission and
instrumentation can be found in \citet{Jansenetal01} and references
therein. The raw data from the EPIC instruments were
processed with the publicly released version of the \xmm\ Science Analysis
System (\textsc{sas v.5.3.3}), using the \textsc{epchain} and
\textsc{emchain} tasks. After filtering for bad pixels and columns, X--ray
events corresponding to patterns 0-12 for the two MOS cameras and patterns
0-4 for the PN camera were accepted.  Investigation of the total count
rates for each field revealed flaring in all three observations. Times when
the total count rate deviated from the mean by more than 3$\sigma$ were
therefore excluded. Effective exposure times after cleaning are given in
Table~\ref{tab:obs}.

Images and spectra were extracted from the cleaned events lists with the
\sas\ task \textsc{evselect}. Point sources were identified using the \sas\ 
sliding-cell detection task \textsc{eboxdetect} to search images for each
camera in five energy bands. The source lists for each band were compared
and combined to produce a final source list for the field. All data within
17\arcs\ of point sources were also removed, excluding the false source
detections associated with the cores of the galaxies. For simple imaging
analysis, the filtering described above was considered sufficient. Event
sets for use in spectral analysis were further cleaned using
\textsc{evselect} with the expression `(FLAG == 0)' to remove all events
potentially contaminated by bad pixels and columns. We allowed the use of
both single and double events in the PN spectra, and single, double, triple
and quadruple events in the MOS spectra. Response files were generated
using the \sas\ tasks \textsc{rmfgen} and \textsc{arfgen}.

Background images and spectra were generated using the ``double
subtraction'' method \citep{Arnaudetal02,Prattetal01}. The ``blank field''
background data sets of \citet{Lumb02} and the ``telescope closed'',
particles only data sets of \citet{Martyetal02} form the basis of this
process. Background images are generated by scaling the ``closed'' data to
match the measured events outside the telescope field of view. The
``blank'' data are then scaled to match the observation exposure time, and a
``soft excess'' image calculated by comparison of this scaled background
to the low energy counts observed in the outer part of the detector field
of view. The source does not extend to the outer part of the field of
view, so the soft excess image should measure only the difference in
background soft emission between the ``blank'' data and that for the
target. The various background components can then be combined to form the
background images. A similar process is used to create background spectra,
again scaling the ``blank'' data to match the observation, and correcting
for differences in the soft background component using a large radius
spectrum.  

\subsection{Chandra observations}
Two of our galaxies, NGC~3585 and NGC~4494 have been observed with
\chandra. A description of the \chandra\ mission and instruments can be
found in \citet{Weisskopfetal02} and references therein. Both datasets were
initially reprocessed using \ciao\ (v2.3), removing bad pixels and
excluding events with \asca\ grades 1, 5 and 7. The data were corrected to
the appropriate gain map, and a background light curve for each was
produced. The light curve for NGC~3585 showed only minimal flaring, and was
cleaned by excluding all times where the count rate deviated from the mean
by more than 3$\sigma$. The effective exposure and other details of this
observation are shown in Table~\ref{tab:obs}.  NGC~4494 unfortunately
showed flaring throughout the observation (on both front and back
illuminated chips), to the extent that removing the flare periods made the
effective exposure so short as to be useless for our purposes. We therefore
did not include this dataset in our analysis.

For NGC~3585, background images and spectra were generated using the blank
sky data described by Markevitch\footnotemark[3]. The data were cleaned to
match the background, and appropriate responses were
created with the \ciao\ tasks \textsc{mkwarf} and \textsc{mkrmf}. As the
ACIS instruments are affected by absorption by material accumulated on the
optical blocking filter, we applied a correction to the responses. When
fitting spectra we generally held the absorption column fixed at the
galactic value of 5.58$\times$10$^{20}$ cm$^{-2}$. Point sources were
identified using the \ciao\ tool \textsc{wavdetect} and were
excluded from the data by removing regions with twice the radius given by
the detection routine.

\footnotetext[3]{http://asc.harvard.edu/cal}

\section{Results}
\label{sec:res}
\begin{figure*}
\epsfig{file=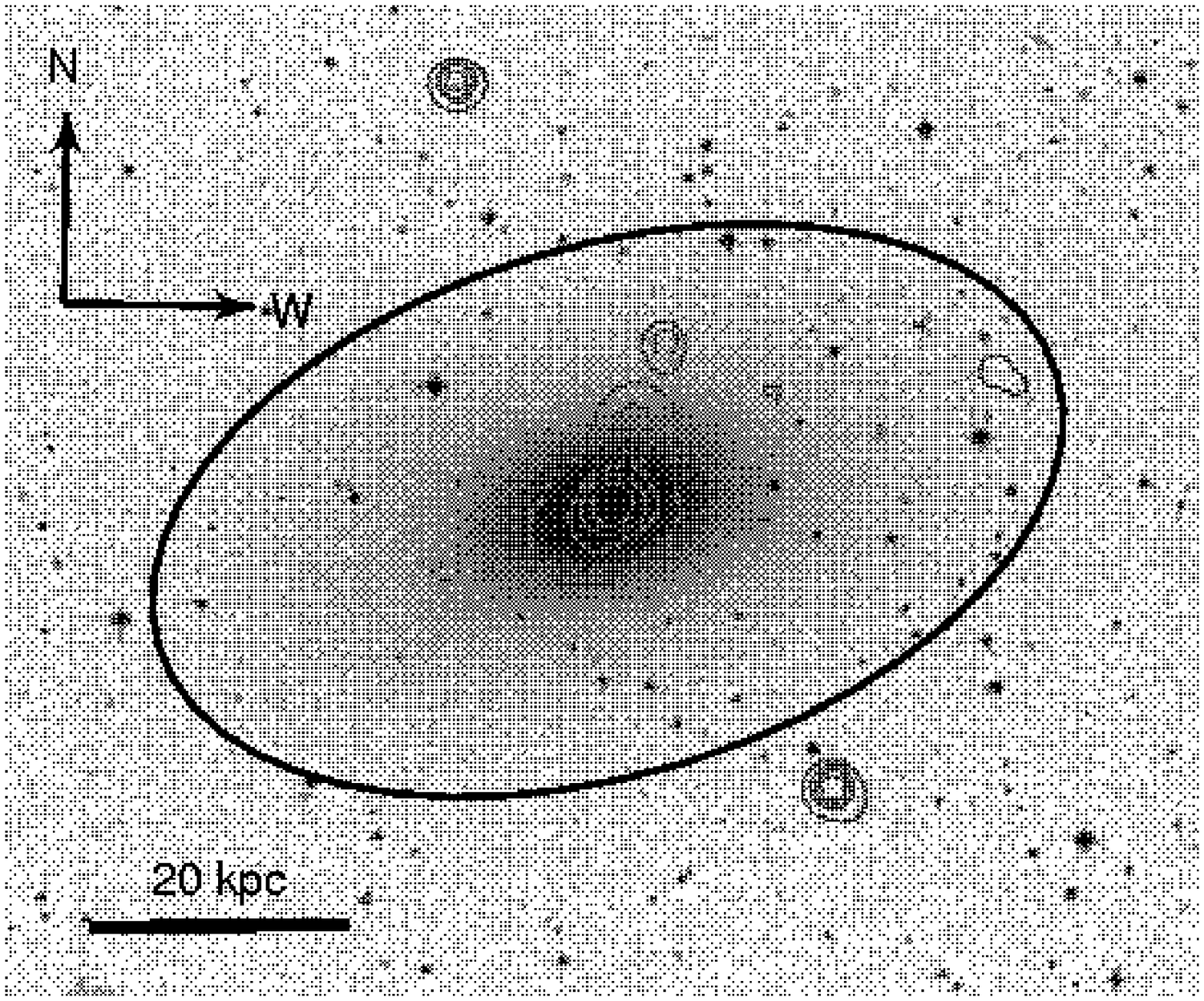,width=5.5cm,bbllx=40,bblly=170,bburx=570,bbury=620,clip=}
\epsfig{file=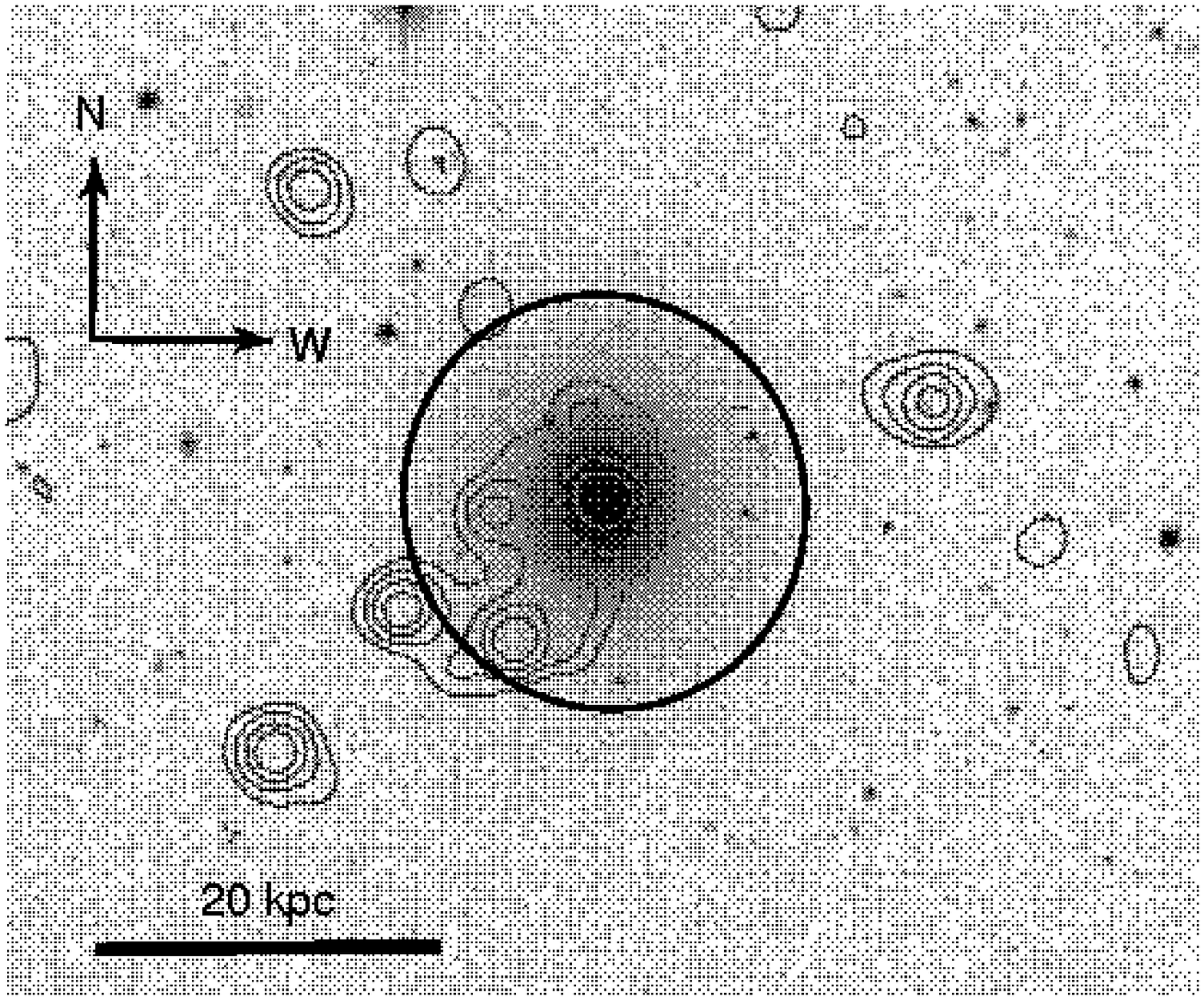,width=5.5cm,bbllx=40,bblly=170,bburx=570,bbury=620,clip=}
\epsfig{file=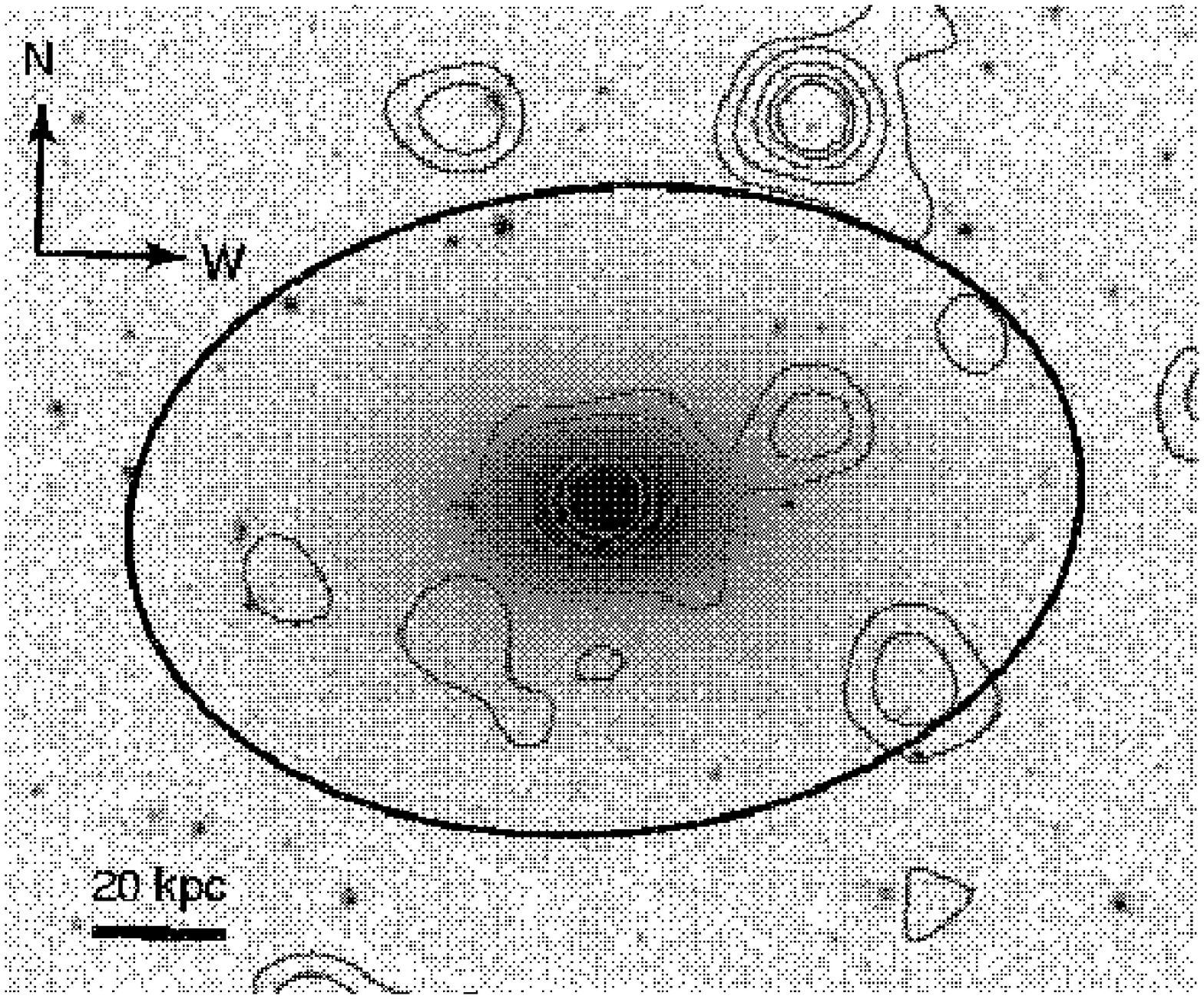,width=5.5cm,bbllx=40,bblly=170,bburx=570,bbury=620,clip=}
\caption{
\label{fig:ovly}
\textit{Left}: NGC~3585, \textit{Centre}: NGC~4494, \textit{Right}:
NGC~5322.  Optical digitized Sky Survey images of the three galaxies with
smoothed \xmm\ X-ray contours superimposed. The large ellipses show the
region enclosed by the \Dtf\ isophote of each galaxy. In each case the peak
X-ray emission coincides with the optical galaxy centre, but asymmetric
extended emission and off-centre point sources are visible. The X-ray
contours are based on mosaiced images from the EPIC cameras, adaptively
smoothed with a signal-to-noise ratio of 10. The 20~kpc scale bars in the
three images correspond to angles of 4.28\arcm\ (NGC~3585), 3.23\arcm\
(NGC~4494) and 2.47\arcm\ (NGC~5322).}
\end{figure*}

We initially prepared adaptively smoothed images of the galaxies from each
dataset. The images from the three \xmm\ EPIC cameras were combined and
smoothed using the \sas\ task \textsc{asmooth} with a maximum
signal-to-noise ratio of 10. \chandra\ data for NGC~3585 was smoothed with
the \ciao\ tool \textsc{csmooth}, using a signal-to-noise range of 3 to 5.
Figure~\ref{fig:ovly} shows X-ray contours derived from the smoothed EPIC
data overlaid on optical images taken from the digitized sky survey.

In all case we found that the X-ray emission from the galaxy was dominated
by point sources. There are extended, diffuse components to the emission,
but they are relatively minor, and none extend out as far as the \Dtf\
radius. There is an apparent lack of correspondence
between the diffuse X-ray structure and the optical appearance of the
galaxies. This can be largely attributed to the effects of point sources
which are unresolved by the EPIC cameras, and in the case of NGC~3585 the
\chandra\ data shows a closer correspondence between diffuse X-ray emission
and the stellar component of the galaxy.

\begin{figure*}
\centerline{\epsfig{file=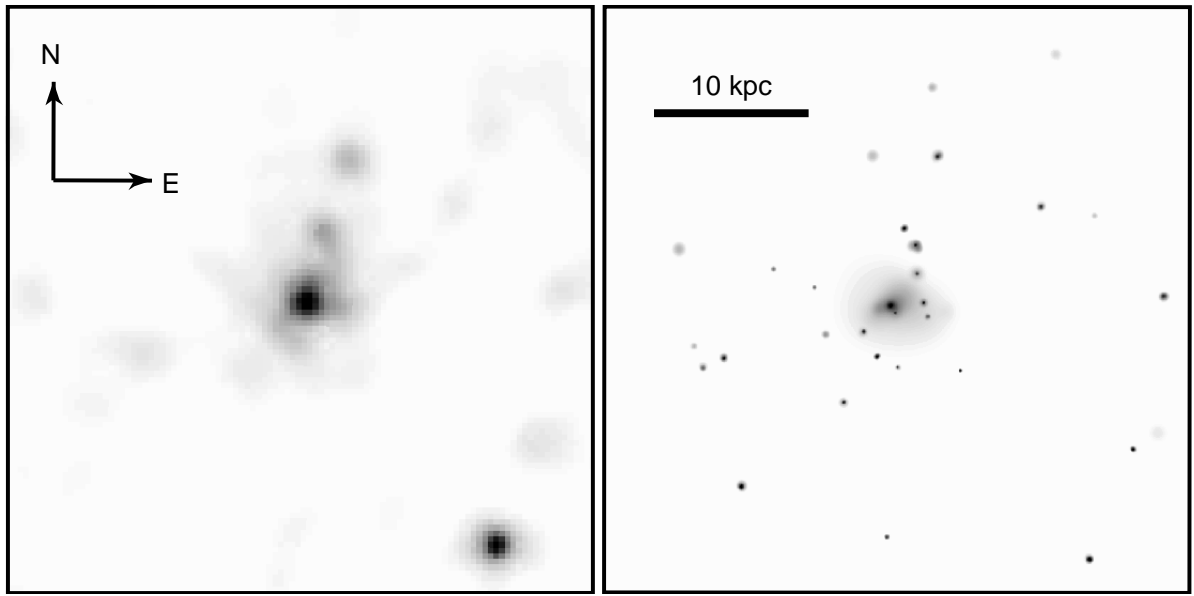,width=16cm}}
\caption{
\label{fig:acismos}
X-ray images of NGC~3585, observed with the XMM EPIC
cameras (\textit{left}) and the Chandra ACIS-S3 chip (\textit{right}).
The scales, positions and roll angles have been matched so that the images
are directly comparable. The 10~kpc scale bar corresponds to an angle of 2.14\arcm. The three EPIC exposures have been combined and
adaptively smoothed using the \sas\ task \textsc{asmooth}, with a maximum
signal-to-noise ratio of 10. The ACIS-S
exposure has been adaptively smoothed with the \ciao\ task
\textsc{csmooth}, with a maximum signal-to-noise ratio of 5. Note that
while some sources can be seen in both images, others are unresolved in the
XMM exposure, and therefore contribute to the diffuse emission.}
\end{figure*}

Figure~\ref{fig:acismos} shows the adaptively smoothed \chandra\ ACIS-S3
image of NGC~3585 alongside the smoothed \xmm\ EPIC. Many of the point
sources visible in the \chandra\ exposure are also visible in the \xmm\
image, but the number of sources which are unresolved by \xmm\ is
clear. Given the small number of counts in each dataset, this contamination
is difficult to deal with. Ideally \chandra\ data would be used to identify
point sources in the field of view, and regions around these would be
removed from the \chandra\ and \xmm\ data before fitting. In practice, the
size of the \xmm\ PSF would force us to omit most of the region containing the
diffuse emission. We therefore identified and removed point sources from each
dataset individually.

\subsection{Spectral fits}
\label{sec:spectral}
For each galaxy we initially extracted spectra from a region of radius
45\arcs, for each of the EPIC cameras. Appropriate background spectra and
responses were generated, and the spectra were simultaneously fitted, using
\textsc{xspec} (v11.2.0). All spectra were grouped to have at least twenty
counts per bin, and fitted using the $\chi^2$ statistic. Power law emission
models were found to give a reasonable fit to the spectra of all three
galaxies above 1~keV, but in all cases we found that the fit below 1~keV
was improved by the introduction of a cool plasma component. We assume this
component is associated with the interstellar medium (ISM) of each galaxy
(we discuss this assumption in Section~\ref{sec:soft}). The results of the
best fitting MEKAL + Power law models are shown in table~\ref{tab:spec},
with power law models for comparison. In all cases the power law model is a
significantly poorer fit, and in the case of NGC~5322 errors on the power
law slope had to be calculated with the \textsc{steppar} command. The
models were fitted in the energy range 0.2-8.0 keV.

\begin{table*}
\begin{center}
\begin{tabular}{lccccccccc}
Galaxy & \NH\ & kT & Abundance & $\Gamma$ & red. $\chi^2$ & d.o.f. & Flux & log \Lx\ & Fraction \\
 & cm$^{-3}$ & (keV) & (\Zsol) & & & & (\ergpcmsqps) & (\ergps) & (PL/MK)\\
\hline\\[-3mm]
NGC 3585 & 5.58$\times$10$^{20}$ & 0.44$^{+0.17}_{-0.11}$ &
0.02$^{+0.03}_{-0.01}$ & 1.23$^{+0.31}_{-0.29}$ & 0.760 & 54 &
1.44$\times$10$^{-13}$ & 39.89 & 0.53/0.47\\[1mm]

 & 5.58$\times$10$^{20}$ & - & - & 1.97$\pm$0.12 & 1.424 & 57 & - & - & - \\

NGC 4494 & 1.56$\times$10$^{20}$ & 0.25$^{+0.14}_{-0.07}$ &
0.02$^{+0.04}_{-0.02}$ & 1.49$^{+0.10}_{-0.11}$ & 1.005 & 153 &
2.34$\times$10$^{-13}$ & 39.86 & 0.88/0.11 \\[1mm]

 & 1.56$\times$10$^{20}$ & - & - & 1.72$\pm$0.05 & 1.153 & 156 & - & - & - \\

NGC 5322 & 1.81$\times$10$^{20}$ & 0.41$^{+0.06}_{-0.05}$ &
0.07$^{+0.03}_{-0.02}$ & 0.88$^{+0.25}_{-0.15}$ & 1.197 & 121 &
2.87$\times$10$^{-13}$ & 40.42 & 0.57/0.43\\[1mm]

 & 1.81$\times$10$^{20}$ & - & - & 2.15$^{+0.05}_{-0.08}$ & 3.15 & 124 & - & - & - \\[1mm]

\hline\\[-3mm]

NGC 3585 (C) & 5.58$\times$10$^{20}$ & 0.37$^{+0.24}_{-0.11}$ &
0.05$^{+0.23}_{-0.04}$ & 1.54$^{+0.47}_{-0.28}$ & 1.182 & 47 & - & - & - \\[1mm]
NGC 3585 (all) & 5.58$\times$10$^{20}$ & 0.41$^{+0.10}_{-0.09}$ &
0.03$\pm$0.01 & 1.39$^{+0.18}_{-0.22}$ & 0.953 & 101 & - & - & - \\
\end{tabular}
\end{center}
\caption{
\label{tab:spec}
Best fit spectral models for our galaxies, with powerlaw fits shown for
comparison. The \xmm\ best fits (and powerlaw only fits) are shown above
the line, the fit to the \chandra\ data (C) for NGC~3585, and to the
combined \chandra\ and \xmm\ data (all). Hydrogen column
was held fixed at the galactic value in all fits. Fluxes are unabsorbed,
and calculated for the 0.2-8.0 keV band. The last column shows the fraction
of flux contributed by the powerlaw/gas components of the model. 90\%
confidence limits are shown for fitted parameters.}
\end{table*}

The need for a low temperature gas component to the fit prompted us to
check several factors which might affect the accuracy of the spectral fit
below 1 keV. The calibration of the EPIC cameras is known to imperfect
below $\sim$0.5~keV, so we carried out the same fits with lower energy
limits of 0.4 and 0.5 keV. In all cases a two component model (power law
with low temperature plasma) was a better fit than a power law model alone.
The parameters of the plasma were poorly constrained, however, and it was
not possible to calculate useful errors on the plasma temperature and
abundance when a minimum temperature of 0.5 keV was used. As our background
data are constructed by a combination of models, including a correction for
galactic soft emission, we also checked to see how well the background
spectra in use match those from a large, source free region of the
observation. As expected they are a good match, particularly in the energy
range of interest (0.2-1.0 keV), suggesting that the low temperature gas
component is not a result of improper background subtraction. We also note
that fits using a power law model with hydrogen column free to vary, while
an improvement on those with \NH\ fixed, do not compare well with the MEKAL
+ power law fits. As a final test, we reanalysed data for NGC~4494 with
\sas\ v5.4.1, a more recent version of the \xmm\ calibration database, and
the more extensive background data files of \citet{ReadPonman03}. The
spectral fitting results were almost identical, the only difference being
slightly smaller error ranges for kT and $\Gamma$.

For NGC~3585 we are able to extract a spectrum for an identical region from
the \chandra\ data. The spectrum was fitted in the energy range 0.3-6.0
keV, the lower limit determined by the \chandra\ calibration and the upper
limit by a lack of counts above 6.0~keV. The results of fitting this data
are shown in the lower portion of Table~\ref{tab:spec}. In general the fit
is quite similar to the fit to the \xmm\ data. Gas temperature, abundance
and the power law photon index are all quite comparable within the errors,
with the main difference being the slightly steeper power law slope in the
\chandra\ fit. The similarity between the two fits is reassuring,
indicating that the low temperature gas component is indeed real. We also
show a simultaneous fit to all four spectra (ACIS-S3 and EPIC MOS 1, MOS 2
and PN), giving slightly better constrained parameters. As a further test
of the fits, we compared the ratio of MEKAL to pwer law emission with the
ratio of counts in point sources (identified by \textsc{wavdetect} in the
\chandra\ data) to counts in diffuse emission. Within 45\arcs, the radius
used for our spectral fits, 54\% of the \chandra\ counts are in point
sources, 46\% in diffuse emission. This compares very well with the results
of the spectral fits (53\%/47\%). For the galaxy as a whole (defined by the
\Dtf\ radius) we find that the ratio is reversed, with 44\% of emission in
point sources and 56\% unresolved or diffuse.

The low temperature component of NGC~5322, while reasonably well modelled
by a single temperature MEKAL model, can also be modelled by a
multi-temperature spectrum. Replacing the MEKAL model with an MKCFLOW
model, designed to represent a steady state cooling flow, improved the fit
significantly (reduced $\chi^2$=1.149 for 120 degrees of freedom).  This
may indicate that the gas in this system is multiphase, although clearly
more detailed spectra would be required before drawing any definite
conclusions. The minimum and maximum temperature of the MKCFLOW model were
kT$_{max}$=0.44$^{+0.12}_{-0.01}$ keV and kT$_{min}$=0.11$^{+0.03}_{-0.11}$
keV. However, it should be noted that these errors were calculated with the
abundance frozen at its best fit value of 0.29 \Zsol. Abundance was poorly
constrained, and when it was allowed to vary the fit was unstable and
\textsc{xspec} was unable to calculate the error bounds. We take this as an
indication that the data do not provide adequate abundance information to
allow an accurate fit with this model. It may suggest that while the
MKCFLOW abundance is consistent with the very low abundances found with
single temperature plasma models, we may be underestimating the abundance
of the gas because we are only able to fit simple models. It has been
demonstrated that fitting a single temperature model to a spectrum from
multi-temperature gas can result in underestimation of abundance
\citep{Buote00,Buote00b}. The power law slope was similar to that found for
the single temperature MEKAL + power law fit.

We also tried fitting each galaxy using a model consisting of a power law,
MEKAL model and a black body model with a temperature kT=50 eV. This third
component was chosen to represent a contribution from unresolved supersoft
sources. \citep{PellegriniFabbiano94} suggested that part of the very soft
component of emission in some early-type galaxies may arise from these
sources, and showed that the \einstein\ data for NGC~4365 were well fit by
a three component model. For our three galaxies, we find that this model
produces only marginal differences from the MEKAL+power law model described
above. The fit for NGC~3585 is slightly worse when the black body component
is included (reduced $\chi^2$=0.785 for 55 degrees of freedom), but the
fits to the other two galaxies are somewhat improved (reduced
$\chi^2$=1.003, 152 d.o.f. for NGC~4494 and reduced $\chi^2$=1.19, 120
d.o.f. for NGC~5322).  The parameters of the model components do not change
significantly, with the exception of the abundance of NGC~4494 which rises
to 0.1$^{+0.28}_{-0.12}$ \Zsol\ (90\% errors). We also carried out fits in
which the black body temperature was increased to 0.1 keV or allowed to
vary freely, but these were less well constrained than the 50 eV model. As
the addition of this model component does not greatly improve the fits, we
choose to continue using the simpler MEKAL+power law model.

Although the numbers of counts in each dataset is low, we were able to
split the \chandra\ data for NGC~3585 and the \xmm\ data for NGC~4494 into
inner and outer spectral regions. The inner region is determined by the
\xmm\ PSF, and is therefore chosen to be a circle of radius 16\arcs. The
outer region is an annulus with outer radius 45\arcs\ and inner radius
enclosing the inner spectral region. The spectrum of each region is not
likely to be detailed enough to provide strong constraints on the
parameters. Our main interest is in determining whether the fraction of
flux contributed by the gas component varies between regions, and whether
it is possible to measure temperature variations in the gas with radius.
The two galaxies show quite different behaviour. NGC~4494 shows very
similar properties in the two regions, with the power law contribution
dominating both. The inner part of NGC~3585 has almost equal flux
contributions from the gas and power law components, but the gas
contribution drops sharply in the outer region. 

One factor which may affect the spectral fits to NGC~4494 is the size of
the \xmm\ PSF. The inner spectral region we have used has a radius
comparable to the half-energy width of the PSF, and the emission is quite
peaked within this region. \citet{Markevitch02} demonstrated that in the
case of A1835, a cluster with a strong peak in core surface brightness,
more than 20\% of the flux in an annulus 15-45\arcm\ from the core
originates from the central 15\arcs\ radius bin. The PSF is also energy
dependent, with low energy photons better focussed by the telescope
mirrors. We must therefore expect that the spectrum of the outer bin of
NGC~4494 contains photons which should be in the inner bin, and that this
contamination will be biased toward higher energy photons. This could
affect the spectral fit, making the outer bin seem to have a harder
spectrum, and disguising a temperature gradient. 

In the case of NGC~3585, the \chandra\ PSF is small enough that this issue
does not arise. We can also compare the ratio of MEKAL to power law
emission to the ratio of counts in detected point sources and in diffuse
emission, as we did for the initial fit to this galaxy. We find that in
both the inner and outer spectral regions, the fraction of counts in
detected point sources is the same, 54\%. This fraction is similar to the
power law emission fraction in the inner bin (51\%), but quite different to
that in the outer bin (83\%). This perhaps suggests that in the outer
spectral bin, a larger part of the X-ray emission is produced by unresolved
point sources, which are spectrally distinct from diffuse hot gas, but
cannot be spatially distinguished from it.

\begin{table}
\begin{center}
\begin{tabular}{lcccc}
Galaxy & \multicolumn{2}{c}{Fraction (PL/MK)} & \multicolumn{2}{c}{kT (keV)} \\
 & Inner & Outer & Inner & Outer \\
\hline\\[-3mm]
NGC 3585 & 0.51/0.49 & 0.83/0.17 & 0.52$^{+0.30}_{-0.16}$ & 0.28$^{+0.23}_{-0.06}$ \\
NGC 4494 & 0.89/0.11 & 0.86/0.14 & 0.19$^{+0.36}_{-0.14}$ & 0.25$^{+0.10}_{-0.08}$ \\
\end{tabular}
\end{center}
\caption{
\label{tab:inout}
Properties of the inner and outer spectral regions of NGC~3585 and
NGC~4494. 90\% errors on temperature are given.
}
\end{table}

%Tcool
Having measured the temperature of the X-ray emitting gas in each galaxy we
can use the model normalistaion to estimate the cooling time of the gas.
Assuming the gas to be approximately isothermal in the 45\arcs\ spectral
fitting region, we can then calculate a mean cooling time based on the gas
temperature and mean density. These values are shown in Table~\ref{tab:SB},
as well as mean gas particle densities for each system. These cooling times
are relatively short, suggesting that either some portion of the gas is
cooling or that there is some source of energy in the galaxies preventing
it from doing so.

\subsection{Surface brightness fits}
Given the relatively low numbers of counts in the datasets for our three
galaxies, we might wish to perform fits to azimuthally averaged surface
brightness profiles. However, at present the \textsc{ciao sherpa} package,
which we use to perform the fits, is not capable of convolving models with
the instrumental PSF in one dimension. In the case of \chandra\ data this
is relatively unimportant, as the PSF is narrow. When fitting \xmm\
data the PSF must be taken into account, particularly when dealing with
objects such as our three galaxies, where the extent of the emission is
only a few times larger than that of the PSF. Regardless of instrument
there is the additional problem that the central point source of each galaxy
may not lie at the same position as the peak of the extended emission. This
could increase the error on a one-dimensional fit, which must by necessity
be centred on one component. We therefore fit all three galaxies in two
dimensions, using images extracted from the cleaned events lists and
performing the fitting using \textsc{sherpa}.

We extracted images and scaled background images of each galaxy, as well as
exposure maps and representative PSFs. Images from the \xmm\ cameras were
binned to 4.4\arcs\ pixels and those from \chandra\ were binned to
1.0\arcs\ pixels. Identified point sources were removed from all images.
Images from the \xmm\ cameras were fit simultaneously, using all three
cameras for each galaxy. We chose to fit images in two energy bands, soft
(0.2-2.0 keV) and hard (3.0-8.0 keV). This was originally intended to allow
a comparison of gas emission with that from unresolved point sources.
However, although the emission from cool gas provides a significant
contribution to the soft band in terms of energy flux, it does not dominate
the photon flux. This means that the images in the soft band are likely to
be strongly affected by the population of unresolved point sources, and
cannot provide an accurate measure of the distribution of cool gas in the
galaxies. A narrower energy band (\eg\ 0.2-0.5 keV) might provide images in
which the gas emission was the main component, but unfortunately we have
insufficient counts to perform fits in this band.

As we had many pixels with few or no counts, we used the Cash statistic
\citep{Cash79} to determine the goodness of the fits. The Cash statistic
does not provide an absolute measure of goodness of fit, but can be used to
measure relative goodness and therefore determine whether a fit is improved
by a particular change in model parameters. As we have no measure of
absolute goodness of fit we determined whether fits were acceptable based
on radial profiles of the data and convolved model (output as images from
\textsc{sherpa}). We emphasize that these azimuthally averaged radial
profiles are not themselves being fitted, but are only inspected to
determine whether the fit is realistic. A further requirement of the Cash
statistic is the use of a background model to describe the background image
in each fit. We chose to use flat background models which have the
advantage that small deviations (`noise') in the background image will not
affect the fit. On the scale over which we are fitting, the background is
relatively (but not perfectly) flat, and the error introduced by using a
flat model should be small.

\begin{figure}
\centerline{\epsfig{file=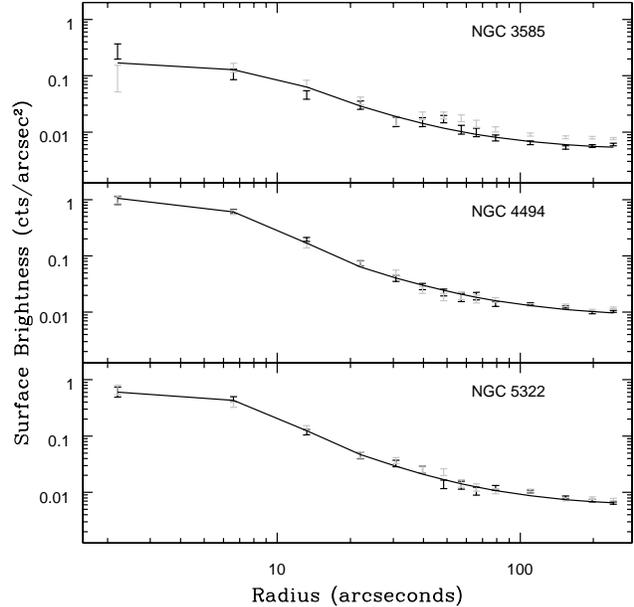,width=9.0cm}}
\caption{
\label{fig:profiles}
Azimuthally averaged radial profiles of the three galaxies in the 0.2-2.0
keV band. The best fitting model is marked by a solid line, and the data
for the MOS 1 \& 2 cameras is marked by black and grey error bars
respectively. PN data points are not shown, as the chip gaps produce
misleading deviations at some radii.
}
\end{figure}

The galaxies were fitted using a combination of beta models, chosen to
represent the extended emission, and where necessary point source models
which represent any central point source contribution. We assume that any
central point source is most likely an AGN. During fitting we found that
the statistics were sufficiently poor that we were unable to calculate
error limits on the parameters for some of the fits. The fits were
unstable, and error fitting in \textsc{sherpa} did not produce reliable
results when tested over multiple error calculations. We were unable to fit any
model to the hard band \xmm\ image of NGC~3585, and so used the \chandra\ 
data instead. This allowed us to model a central point source but we did
not find any extended component. The soft band fits to the \xmm\ and
\chandra\ data for NGC~3585 were similar.  Table~\ref{tab:SB} shows the
results of the fits, and Figure~\ref{fig:profiles} shows azimuthally
averaged radial profiles of the data and best fitting model for each
galaxy, in the soft band.

\begin{table*}
\begin{center}
\begin{tabular}{lcccccccc}
Galaxy & Instrument & Band & r$_{core}$ & \Bfit\ & Point Source? &
r$_{max}$ & t$_{cool}$ & n \\
 & & (keV) & (pc) & & & (kpc) & (Myr) & (cm$^{-3}$) \\
\hline\\[-3mm]
NGC 3585 & EPIC & 0.2-2.0 & 29.1$^{+6.7}_{-11.3}$ & 0.45$^{+0.01}_{-0.01}$ & Y & 7.01 & 639.6 & 0.014 \\
NGC 3585 & ACIS-S3 & 3.0-8.0 & \multicolumn{2}{c}{Point Source only} & Y & & & \\
NGC 4494 & EPIC & 0.2-2.0 & 84.1$^{+13.9}_{-12.8}$ & 0.42$^{+0.01}_{-0.01}$ & Y & 6.96 & 1863.9 & 0.017 \\
NGC 4494 & EPIC & 3.0-8.0 & 431.8 & 0.65 & N & & &\\
NGC 5322 & EPIC & 0.2-2.0 & 139.2 & 0.43 & Y & 12.14 & 825.5 & 0.009 \\
NGC 5322 & EPIC & 3.0-8.0 & \multicolumn{2}{c}{Point Source only} & Y & & &\\
\end{tabular}
\end{center}
\caption{
\label{tab:SB}
Surface brightness fits to images of our galaxies in different energy
bands. All errors are 1$\sigma$ confidence limits. Values given without
errors indicate that the error calculation failed owing to poor
statistics. The point source component is always central, and may
correspond to AGN in the cores of the galaxies. r$_{max}$ is the radius to
which we can detect galaxy emission at 3$\sigma$ confidence above the
background using all three EPIC cameras. Mean gas density and cooling time
are derived using the best fit mean temperatures from the spectral fits to
each galaxy.
}
\end{table*}

Although the radial profiles show the fits to be fairly good
representations of the data, Some differences between model and data are
visible. The profile for NGC~3585 suggests that the model is
underestimating the surface brightness at radii of 40-80\arcs. This is
likely caused by the apparent extension of the galaxy emission to the
northwest, visible in the left hand panel of Figure~\ref{fig:acismos}. As
the right hand panel of that figure shows, this is largely caused by
point sources which are poorly resolved by \xmm\ but which are clear in the
\chandra\ data. We note that the surface brightness fit to the \chandra\
data is similar to that we find for \xmm, suggesting that this problem has
not significantly biased the fit.

The formal 1$\sigma$ errors on \Bfit, which are calculated using the
\textsc{projection} task in \textsc{sherpa}, are rather small given the
apparent uncertainties in the data and the difficulty of finding stable
fits. However, using alternate methods of error calculation in
\textsc{sherpa} produced similar error estimates, suggesting that either
these errors are accurate or that there is a general problem with the
calculation of the error on the \Bfit\ parameter in this software package.
The core radii of the king models are all comparable to the pixel scale of
the images used for fitting. This may indicate the presence of central
point sources even where these are not required for the fit. One clear
trend does emerge from these fits however; the extended emission is
primarily seen in the soft band, and has a very flat profile, flatter than
that which would be expected for stars. A de~Vaucouleurs profile typical of
elliptical galaxies is similar to a beta model with \Bfit=0.5
\citep{Brownbreg01}. The flatness of the profile could indicate that the
gas in the galaxies is more extended than the stellar population. However,
unresolved point sources in globular clusters around the galaxies could
also be responsible. In NGC~4494, where we have sufficient numbers of
counts in the hard band to detect the extended component, the model fit
shows a steeper slope than in the soft band. This could be compared to the
change in temperature with radius seen in NGC~3585, but given the lack of
such a change in NGC~4494, it is perhaps more likely a product of the lack
of counts in the hard band. We were unable to calculate the errors on the
slope in the hard band, indicating that they are probably large, and likely
overlap the slope found in the soft band.

\section{Discussion}
\label{sec:disc}
\begin{figure}
\centerline{\epsfig{file=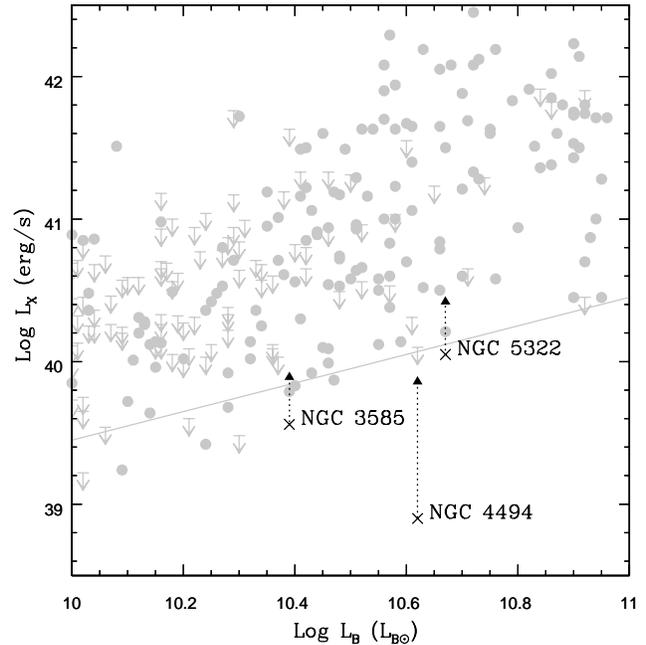,width=9cm,bbllx=20,bblly=210,bburx=562,bbury=779,clip=}}
\caption{
\label{fig:lxlb}
A plot of log \Lx\ against log \LB\ for early type galaxies. Points in grey
are drawn from the catalogue of \protect\citet{OFP01cat}, with circles
designating detections and arrows upper limits. Points for the three
galaxies in this work are marked in black, triangles showing the total
luminosity, crosses the luminosity of the gas component. The optical
luminosities for the galaxies are taken from \protect\citet{OFP01cat}. The
solid grey line marks an estimate of the mean contribution to galaxy X-ray
luminosity from discrete sources \protect\citep{Ciottietal91}.  }
\end{figure}

The most important factor arising from the X-ray observations of these
galaxies is the confirmation of how unusually X-ray faint they are.
Figure~\ref{fig:lxlb} shows X-ray luminosity plotted against B-band optical
luminosity for a range of early-type galaxies, with our three galaxies
shown for comparison. The general population of galaxies forms a band on
this plot, the lower limit of which corresponds roughly to the estimate of
X-ray emission from discrete sources from \citet{Ciottietal91}. This makes
sense, in that galaxies in the lower part of the band are presumably
dominated by point sources, with little or no gas emission, while those
with large gas halos fall in the upper part of the band. The three galaxies
we are considering all fall near the lower limit of the band, NGC~4494 in
fact falling below the edge of the band. When we consider only the
luminosity from the soft spectral component, all three fall further, with
NGC~4494 perhaps two orders of magnitude fainter than galaxies with similar
optical luminosities.

\begin{figure}
\centerline{\epsfig{file=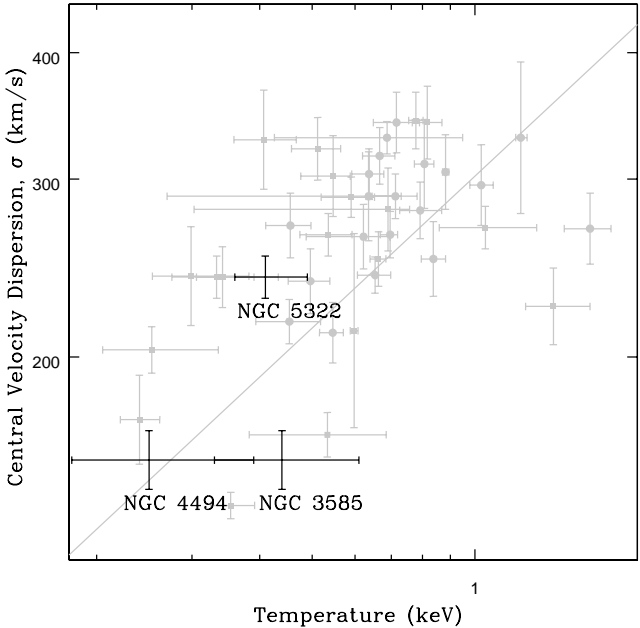,width=9cm,bbllx=20,bblly=210,bburx=562,bbury=779,clip=}}
\caption{
\label{fig:sigmaT}
A plot of stellar velocity dispersion against X-ray temperature for early
type galaxies. Points in grey are drawn from the \rosat\ sample of
\protect\citet{OPC03}, which consists primarily of large, gas rich
ellipticals. Error bars for these points show 1$\sigma$ confidence regions.
Points for the three galaxies in this work are marked in black, with
1$\sigma$ errors on velocity dispersion and 90\% errors on temperature. The
solid grey line marks \Bspec=1, equipartition of energy between the stars
and gas in the galaxies.}
\end{figure}

A further comparison can be seen in Figure~\ref{fig:sigmaT}, a plot of
stellar velocity dispersion against X-ray temperature. We have compared the
three galaxies from this work with the galaxies in the sample of
\citet{OPC03}, which is mainly comprised of large ellipticals in the
centres of galaxy groups. The sample does have a few less massive objects
however, and it is notable that NGC~4494 and NGC~3585 fall in the bottom
left hand corner of the plot, close to points for NGC~5128 (which hosts
Centaurus~A and has only a moderate amount of X-ray emitting gas),
NGC~1549, NGC~1553 \citep[an elliptical and an S0 galaxy which are
interacting,][]{MalinCarter83}, and NGC~4697 \citep[another gas-poor
elliptical,][]{SarazinIrBreg01}. All three galaxies are fairly close to the
line \Bspec=1, and NGC~4494 is consistent with the line. \Bspec\ is defined
as

\begin{equation}
\beta_{spec} = \frac{\mu m_p \sigma^2}{kT},
\end{equation}

\noindent where $\sigma$ is the stellar velocity dispersion, $\mu$ is the
mean mass per particle and m$_p$ is the proton mass.  A value of \Bspec=1
indicates that there is energy equipartition between X-ray gas and stellar
velocity dispersion. This may suggest that the main source of the gas is
stellar mass loss, so that its current temperature is the product of
thermalisation of the velocity it had when ejected from the stars. The fact
that there is no major difference between our three low X-ray luminosity
galaxies and those in the \citet{OPC03} sample suggests that the source of
the gas and the means by which it is heated may be similar.

\subsection{The soft spectral component}
\label{sec:soft}
An important question for these galaxies is the origin of the soft spectral
component, which we have modelled as a plasma with temperatures of
0.25-0.44 keV. As discussed in Section~\ref{sec:spectral}, this component
does not appear to be a product of poor background subtraction, and is
found in both spectral regions used for NGC~3585 and NGC~4494. Although
this indicates that the emission is extended to some extent, we are not
able to confirm whether it arises from a diffuse gas component or from
unresolved point source emission. As the soft band surface brightness fits
are affected by emission from the X-ray binary component, these cannot
provide any information on the distribution of the soft spectral component. 

This issue is not new. Observations of low X-ray luminosity ellipticals
with \einstein\ showed that they had a very soft emission component with a
temperature of $\sim$0.2-0.3 keV \citep{PellegriniFabbiano94}. In one
elliptical galaxy, NGC~4365, it was shown that the \einstein\ data could be
modelled using a combination of components representing hard emission from
X-ray binaries, a 0.6 keV plasma, and soft emission from stellar sources
whose mean temperature was $\sim$0.1 keV. Sources such as M dwarf stars, RS
Canum Venaticorum (RS CVn) late-type stellar binaries and the so-called
``supersoft'' X-ray sources were suggested as possible contributors to the
softest emission component, but it was shown that they could only produce
the required luminosity and temperature when combined. Any of these three
classes of source was unlikely to be the sole source of the soft component,
and a gas component was still required to model the galaxy.

More recent studies of ellipticals using \chandra\ have resolved a small
number of ``supersoft'' sources \citep[\textit{e.g.}][]{SarazinIrBreg01},
and studies of sources within the local group have shown that they are
characterised by spectra with temperatures kT$<$90 eV and luminosities of
$\sim$10$^{36}$ to $\sim$10$^{38}$ \ergps\ \citep{Kahabka02}. We note that
these temperatures are substantially lower than those we find for the soft
emission in our galaxies, even if we adopt the \citet{PellegriniFabbiano94}
model and increase the temperature to $\sim$0.1 keV. However, these sources
may contribute some of the soft emission, and we therefore investigate the
possibility.

We have attempted to fit our galaxies with models including a soft
component such as that produced by supersoft sources (a 50 eV black body, see
Section~\ref{sec:spectral}) but this model is not a significant improvement
over a MEKAL+power law model, and is worse in the case of NGC~3585. Fixing
the temperature at 0.1 keV, the temperature found by
\citet{PellegriniFabbiano94}, does not improve the fit significantly and
increases the error bounds on the parameters. If the temperature of this
supersoft component is left free, the fits become unstable and poorly
constrained. If we assume the model with the 50 eV black body component is
accurate, we find that in NGC~4494 the black body component contributes a
similar energy flux to the MEKAL component (8 per cent of the total
emission), while in NGC~5322 the MEKAL component is $\sim$4.5 times as
luminous as the black body. 

These results suggest that it is possible that a third, very soft component
such as that described by \citet{PellegriniFabbiano94} may contribute to
the emission in our galaxies, to a varying degree. The poor fit to the data
for NGC~3585 argues against such a component making any significant
contribution, but in NGC~5322 a small contribution is possible. In
NGC~4494, the contribution may equal that of the gas component. However,
these results should be considered as upper limits, as the spectral fits do
not show that an additional black body component improves the fits
significantly. The additional component also has very little effect on the
properties of the other spectral components, except the increase in
abundance for NGC~4494. It therefore seems reasonable to continue with the
assumption that the soft spectral component represents emission from hot
gas, with the caveat that the normalisation may be slightly overestimated.

\subsection{Stellar properties}
All three galaxies have optical surface brightness profiles which are
sharply peaked in the core \citep{Michard98}. It has been suggested that
this type of profile is associated with disky, rapidly rotating galaxies
\citep{Nietoetal91} with low X-ray and radio luminosities
\citep{Benderetal89}. The differences between galaxies with this structure
and those with boxy isophotes, slow rotational velocities and flat central
surface brightness distributions may be an indication of different origins.
In this model disky ellipticals are formed through dissipational collapse,
or perhaps a collapse followed by only minor mergers, while boxy
ellipticals with high X-ray luminosities are the product of major mergers.

However, the galaxies in this sample have structures which may indicate
that they have undergone major mergers in the past. NGC~3585 is a disky
elliptical, containing a rapidly rotating disk component embedded within a
non-rotating bulge \citep{ScorzaBender95}. The difference in rotation
velocities (280 \kmps\ for the disk, 45 \kmps\ for the bulge) indicates
different origins for the two components, and a probably merger origin for
the galaxy. NGC~5322 has a counter-rotating core \citep{Bender88}, and is
boxy in its outer regions, only becoming disky in the core. It can be
modelled by assuming that the main body is a slowly rotating (30 \kmps) boxy
ellipsoid with a small disk in the core \citep{ScorzaBender95}. Such a disk
could be formed by a centrally concentrated burst of star formation after a
merger or strong interaction, or could be a remnant of a progenitor galaxy.
In either case, a merger origin is indicated for NGC~5322. NGC~4494
contains a small dust disk \citep{Tranetal01} and a central stellar disk
\citep{Carolloetal97}. The galaxy core is kinematically distinct, but the
stellar disk is not more metal poor than the rest of the galaxy,
suggesting that it formed through dissipative collapse during a merger
between gas-rich galaxies. All three galaxies have globular cluster
populations with a bimodal colour distribution \citep{vandenBergh01}.
NGC~4494 and NGC~5322 have fairly low 'local' specific frequencies of
globular clusters \citep{Kundu99}. Based on the distribution and colours of
globular clusters in these two objects, studied as part of a sample of
galaxies with kinematically distinct cores, \citet{Forbesetal96} conclude
that the most likely model for the formation of these galaxies is through
merger of massive gas-rich progenitors at an early epoch.

Given the optical evidence of formation through merger for these galaxies
and their low X-ray luminosity, it seems likely that they are relatively
young objects, only a few Gyr from formation through merger. Unfortunately
only NGC~3585 has a spectroscopic age (the age of the stellar population,
based on fitting stellar evolution models to optical spectra), but this
confirms it to be young, with an age of $\sim$3 Gyr
\citep{Terlevichforbes00}.  \citet{Schweizerseitzers92} describe the age of
NGC~5322 through their fine structure parameter, in which values are
assigned to galaxies based on the degree of disturbance (shells, ripples,
tidal tails, \etc) visible in their structure. They find a value of
$\Sigma$=2, indicating moderate disturbance, and supporting the assumption
of a young age for the galaxy, as such features generally fade over the
course of a few Gyr. It is unfortunate that all three galaxies do not have
well determined spectroscopic ages, but it seems clear that two of these
objects at least are relatively young.

\subsection{Dark matter and halo stripping}
In Section~\ref{sec:intro} we described three possible reasons why these
galaxies might be X-ray under-luminous. The first of these was that the
galaxies might have minimal dark matter halos, leaving them with too little
mass to retain an X-ray halo. In the case of NGC~4494, there is some
evidence to suggest that this may be the case. \citet{Romanowskyetal03} use
planetary nebulae to trace the velocity dispersion profile of the galaxy to
\gtsim3 $R_e$ ($\sim$150\arcs\ or $\sim$15.2 kpc) and show that within this
region the stellar population alone is capable of producing the observed
profile. This suggests that either the system has a minimal (or no) dark
halo, or that the halo does not show the central concentration expected
from cold dark matter formation models. In either case, we must expect that
the galaxy's ability to retain hot gas would be greatly reduced by this
lack of a centrally concentrated dark halo. Indeed, the lack of a bright
X-ray halo could be considered as evidence that the system does not have a
significant dark matter halo. We must also consider the
possibility that our other galaxies have a similar lack of dark matter. It
is worth noting that the mass-to-light ratios of NGC~4494 and
NGC~5322 determined from stellar velocity dispersion measurements, are
slightly lower than average \citep{Baconetal85}. The two galaxies were
analysed as part of a sample with a mean mass-to-light ratio of $\sim$10,
and were found to have ratios of 8.3$\pm$0.16 (NGC~4494) and 9.0$\pm$0.43
(NGC~5322).

Perhaps the most likely way to produce a galaxy with a lower than average
quantity of dark matter would be to strip the dark matter halo. Such tidal
stripping could also remove X-ray gas and globular clusters. Stripping
would happen during interactions between the galaxy and its neighbours, but
this seems unlikely in the case of our objects.  NGC~3585 and NGC~5322 are
the dominant galaxies of small groups \citep{Garcia93}, an environment
which would seem more likely to enhance their halos than reduce them.
NGC~4494 lies in a small group in the outskirts of the Virgo cluster, the
other members being gas-rich spiral galaxies. In each case, the galaxies we
have observed are the most massive elliptical in the group, and it seems
more plausible that they would strip matter from their neighbours rather
than the other way around. Simulations of interactions between galaxies in
compact groups have shown that some of the dark matter originally
associated with the member galaxies can be stripped and left as a diffuse
group halo \citep{Barnes89}. These simulations also show that the dominant
(elliptical) merger remnant does possess its own dark halo, but there could
be circumstances in which this does not hold. It is very unfortunate that
these observations are not sufficiently deep to allow measurement of the
surface brightness profile of the gas in each galaxy, as this would allow
us to estimate the total mass profile of the galaxies. This is clearly an
area in which a larger sample of ellipticals with velocity dispersion
measurements extending to large radii as well as high quality X-ray data is
needed.

We can more confidently dismiss the option of ram-pressure stripping. A
galaxy, even a relatively massive one, could be partially stripped of its
X-ray halo by passage through the dense core of a galaxy cluster
\citep{Acremanetal03}. However, the position of our galaxies in small
groups argues against such an event in their history. The galaxies have
small velocities relative to the other members of the group, and neither
our \xmm\ and \chandra\ data nor our analysis of the archival \rosat\ 
observations \citep{OFP01cat} show evidence of large group halos of X-ray
emitting gas. Based on the background we measure immediately around our
target galaxies, we estimated 3$\sigma$ upper limits on the density of gas
in each group potential well, and also upper limits on the luminosity of
each group. We assumed the halos to have temperatures of 0.3 keV, 0.3
\Zsol\ abundance, and to extend to a radius of 100 kpc. The chosen values
are low compared to many bright groups \citep{Helsdonponman00} but are
probably realistic for low mass, low luminosity systems. These upper limits
are given in Table~\ref{tab:limits}, and show that the intra-group medium
is of relatively low density. At velocities of (at most) a few hundred
\kmps\ such diffuse gas is very unlikely to have affected our target
galaxies. The presence of gas-rich spiral galaxies in the groups
demonstrates clearly that no ram-pressure stripping has happened to the
group as a whole. We cannot therefore explain the lack of dark matter and
hot gas in NGC~4494 by postulating a passage through the core of Virgo.

\begin{table}
\begin{center}
\begin{tabular}{l|cc}
Galaxy & n & log \Lx\ \\
 & (10$^{-5}$ cm$^{-3}$) & (\ergps) \\
\hline
NGC 3585 & 24.51 & 41.88 \\
NGC 4494 & 4.44 & 39.72 \\
NGC 5322 & 6.86 & 41.14 \\
\end{tabular}
\end{center}
\caption{
\label{tab:limits} Estimated 3$\sigma$ upper limits on the density and
luminosity of undetected group X-ray halos surrounding our target
galaxies. We assume the group gas to have temperatures of 0.3 keV and 0.3
\Zsol\ abundance. Density limits are based on the counts required for a
3$\sigma$ detection in an annulus whose inner radius is the radius to which
we detect galaxy emission and of 3\arcm\ width. Luminosity limits assume
the halo to extend to 100 kpc around the target galaxies.}
\end{table}

\subsection{X-ray halo formation models}

The third option mentioned in Section~\ref{sec:intro} is that these are
simply young galaxies.  Numerical simulations of the formation of the large
X-ray luminous gas halos associated with early-type galaxies suggest that
the process is affected by a number of factors, including galaxy mass
(stellar and dark matter), supernova rate and environment. For the galaxies
we have observed, which although massive are not surrounded by a dense
intergalactic medium, models such as those of \citet{Ciottietal91},
\citet{Pellegriniciotti98} or \citet{DavidForJo91} may be the most
applicable. These `galaxy wind' models all pass through three main phases,
characterised by the movement of gas in the ISM. The phases can be
described as (1) a supersonic outflow driven by the high supernova rate
shortly after galaxy formation, (2) a slower sub-sonic outflow once the
supernova rate drops, with consequent higher gas densities, and (3) a
sub-sonic inflow driven by cooling of the densest gas in the core of the
galaxy. The models of \citet{Pellegriniciotti98} are slightly different in
that they produce `partial winds' in which the inner portion of the galaxy
hosts a cooling flow while the outer part is still driving an outflowing
wind. In this case, the radius inside which the gas flows inwards expands
with galaxy age. \citet{DavidForJo91} give details of models both with and
without dark matter halos. In those without dark matter, the galaxies never
reach phase (3), and at most develop partial inflows. For galaxies of
similar \LB\ to ours, the model predicts supersonic winds at both early and
late stages, with a sub-sonic partial wind at intermediate age.  In all of
the models, the mass of the system and the supernova rate are the major
factors in determining how long the galaxy remains in a particular phase.
Low mass galaxies may not have passed out of the supersonic wind phase by
the present epoch, and galaxies which have undergone major mergers in the
recent past may still be in the process of regenerating their halos. Our
galaxies may fall in to either category; if they lack sufficient dark
matter they may be unable to retain a significant gaseous halo, and if they
do have massive dark halos they may be too young to have built up a large
quantity of X-ray gas.

The galaxy wind models are supported to some extent by observation. Total
X-ray luminosity and \LxLb\ ratio are generally low for dynamically young
early-type galaxies \citep{OFP01age,Sansom00,MackieFabb96}, particularly
for those galaxies whose stellar population has a spectroscopically
determined mean age of $<$5 Gyr. This timescale compares well with the time
predicted by the models for the buildup of a halo. The models suggest that
the X-ray luminosity at later times will depend on the development of the
gas flow, with lower mass systems remaining fairly faint while more massive
objects become highly luminous on a fairly short timescale (1-3 Gyr). The
X-ray observations show that the mean X-ray luminosity of a sample of
galaxies rises slowly over a long timescale \citep{OFP01age}. This may be
an indication that the increase in luminosity is somewhat slower
than predicted by the models. In any case, the low luminosity of our
three galaxies suggests that they may be relatively young objects.

\begin{figure}
\centerline{\epsfig{file=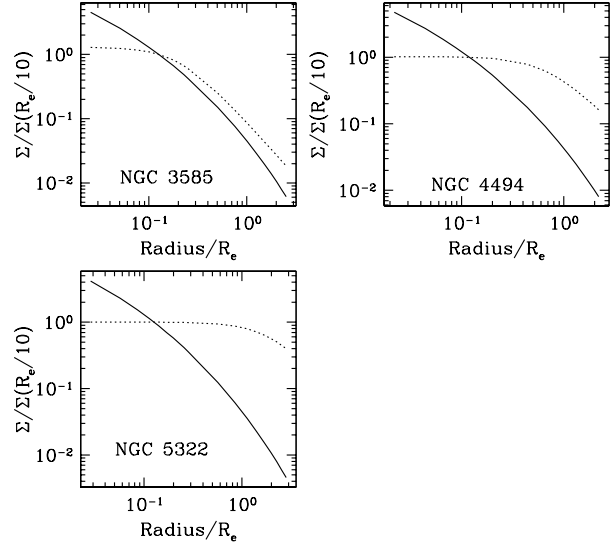,width=8.5cm}}
\caption{
\label{fig:optical}
Optical surface brightness profiles for the three galaxies, marked by solid
lines, compared to the X-ray profile of the gas component, marked by dotted
lines. Effective radii ($R_e$) are taken from \protect\citet{Faberetal89}.}
\end{figure}

The models make numerous predictions about factors which we can measure,
such as the gas density profile, temperature, and the mass of gas available
from stellar mass loss. One of the most important concerns the relative
gradient of the X-ray surface brightness (or gas density) profile compared
to the optical profile. Both the Ciotti \etal\ and David \etal\ models
predict that the X-ray profile will be steeper or more centrally
concentrated than the optical during the supersonic outflow and inflow
phases, but flatter than the optical during the sub-sonic outflow phase.
The Pellegrini \& Ciotti models tend to be steeper in all phases, as the
central cooling inflow adds a peak to the surface brightness profile. Our
surface brightness profiles are probably strongly influenced by emission
from unresolved point sources, and so cannot be used for a direct
comparison. However, any strong correlation (or lack thereof) might still
be of some interest, as the gas does have some influence on the profile.
Figure~\ref{fig:optical} shows r$^{1/4}$ law profiles for each of our
galaxies compared with the X-ray surface brightness profiles. In all cases
the X-ray profile is flatter than the optical, though this is more
pronounced in NGC~4494 and NGC~5322, where the core radius is considerably
larger than in NGC~3585. If this difference is produced by extended gas
emission rather than unresolved point source emission in the globular
cluster population, it suggests that if the galaxy wind models are correct,
the galaxies are in a sub-sonic outflow phase, a surprising result given
their low luminosity. However, we emphasize that this result is biased by
the unresolved point source emission and cannot be regarded as definitive.

As well as predicting the slope of the density profile, the stellar wind
models make predictions about the absolute density of the gas. The models
also make predictions about other factors, such as the mass of gas
available from stellar mass loss and the gas temperature profile. Again,
the mass and density of the gas depend on the age of the system and its
current wind phase; if the galaxies are in a supersonic wind phase we
expect a large stellar mass loss rate, but almost all of this mass is lost
to the galaxy. Our spectral fits suggest that the three galaxies have gas
masses of $\sim$10$^6$ \Msol\ if the gas is evenly distributed. If we
assume a density profile for the gas based on the surface brightness
profile and normalised using the spectral fits, these masses rise to 
2$\times$10$^7$-10$^8$ \Msol\ and we would predict peak gas
densities of $\sim$0.1 cm$^{-3}$. Assuming our galaxies to be relatively
young objects with ages $<$5 Gyr, the models of Ciotti \etal\ and
Pellegrini \& Ciotti generally over-predict the amount of gas in the
system, suggesting that we should see masses $>$10$^9$ \Msol. The exception
may be models in which the galaxy is rapidly rotating, as the gas in these
models gains more energy from the stellar motions and is more easily lost
to the galaxy. The David \etal\ models do a better job of predicting the
gas mass, with values of 10$^7$-10$^8$ \Msol, but under-predict the central
density by a factor of $\sim$10. This is the case for models both with and
without dark matter. The Pellegrini \& Ciotti models match our observed
temperatures relatively well, predicting kT$\sim$0.48 keV, with a decline
in temperature with radius which would agree with that we observe in
NGC~3585. The David \etal\ models predict a slightly higher temperature,
$\sim$0.7 keV, and the Ciotti \etal\ models over-predict kT by a factor of
$\sim$3-4.

\subsubsection{Abundance}
However, the most important problem for all the models is the abundance of
the gas. All the models assume that the gas is produced through stellar
mass loss and must therefore be enriched. Ciotti \etal\ state that their
models predict winds with metallicities several times solar, which would be
difficult to reconcile with the very sub-solar abundances we observe. As the
low numbers of counts in our spectra lead to rather poorly constrained
models, we have carried out simulations to determine whether multi-phase
gas models or the contributions from X-ray binaries could cause us to
underestimate the gas abundance by a significant amount. In particular we
wanted to determine whether the abundances we measured might be reduced
through the ``Fe bias'' effect described by \citet{Buote00b}, which is most
influential in low temperature systems such as groups and galaxies.

We used
\textsc{xspec} to simulate spectra for NGC~4494 (the galaxy for which we
have the best statistics), initially assuming a gas component with an
abundance of 0.3 \Zsol\ and a temperature as measured from the observed
spectra of the galaxy. We modelled the X-ray binary contribution using a
powerlaw component with a photon index of $\Gamma$=1.56, which
\citet{IrAtheyBreg03} have shown to be a good representation of the binary
population in a range of early-type galaxies. The relative contributions of
the two components were then varied, keeping the total number of detected
counts at $\sim$4500. We found that if the powerlaw component is dominant,
contributing 90 per cent or more of the counts, the measured abundance can
be reduced by a factor of $\sim$2. However, abundances less than 0.3 \Zsol\
are required to reproduce the results we observe, and in these cases it is
clear that the powerlaw is the dominant component. This does not agree with
what we observe in the three galaxies. Similarly, we simulated galaxies in
which the gas emission was produced by gas at two temperatures, one held at
the measured temperature, the other varied between 0.2 and 1 keV. Abundance
was held to be the same for both components and varied between 0.1 and 1
\Zsol. Again we found that it was difficult to disguise the true abundance
of the gas, and that low abundances ($<$0.15 \Zsol) were required. These
tests strongly suggest that while the abundances we measure may be
underestimated, they cannot be as high as those predicted by the galaxy
wind models.

There are numerous examples of galaxies observed with a range of satellites
which show exceptionally sub-solar abundances
\citep[\textit{e.g.}][]{Daviswhite96,Matsumotoetal97,IrwinSarBreg02}. The
difference between these abundances and the optically determined abundances
of the stellar populations has been the subject of debate for a number of
years. In the case of \rosat\ the spectral resolution of the PSPC camera
made modelling of the abundance rather unreliable, and for \asca\ 
observations, it has been shown that most ellipticals can be accurately
modelled using multi-temperature models with near solar abundances
\citep{Matsushita00}. More recent studies have also shown that bright
ellipticals have abundances more consistent with the stellar population
when modelled with multi-temperature models
\citep{Buoteetal03b,Matsushitaetal03}. However, the problem appears to
persist for galaxies with low X-ray gas content, as several recent studies
show very low abundances similar to those we observe in our targets
\citep{SarazinIrBreg01,IrwinSarBreg02,Kraftetal03}. Given the relatively
small numbers of counts detected from the ISM of these galaxies, it is
still possible that the abundance could be underestimated owing to the use
of overly simplistic models. However, our spectral simulations suggest that
this is not the case for our galaxies. It has also been noted
\citep{FinoguenovJones00} that in many
studies the angular scales on which stellar and ISM abundance are
determined are different; optical metallicities tend to be measured in the
core of each galaxy using apertures considerably smaller than those
available to X-ray satellites such as \rosat\ or \asca. However, the advent
of \chandra\ has altered this situation to the point where abundances can
be measured on very similar scales for the ISM and stellar
population. \citet{IrwinSarBreg02} demonstrate that even in the very center
of NGC~1291 the ISM has very low abundance. Our galaxies, in which
abundance is determined on a rather coarser scale, support the suggestion
that low X-ray luminosity and low abundance are linked. This is clearly a
problem for the galaxy wind models.

\section{Conclusions}
There appear to be two main possible explanations for the low X-ray
luminosity and low gas content of the three galaxies we have observed. The
galaxies may have minimal or nonexistent dark matter halos, reducing their
ability to retain gas. They may be young objects, in which the energy
released by supernovae is still great enough to drive low density
outflowing stellar winds, effectively removing gas from the galaxies. The
available optical data suggests that both of these factors may be at work,
combining to leave only the small amounts of gas which we detect. Several
important issues arise from these results. Firstly, if the galaxies do lack
massive dark matter halos, why is this so? Stripping of the dark matter
seems unlikely given the environment of the systems, and current galaxy
formation models predict that cold dark matter halos are a fundamental
feature of early-type galaxies. Secondly, the galaxy wind models do suggest
that enough gas should be available to produce the emission we see, but we
do not find a close match between their predictions about gas density and
the density profiles we measure. Thirdly, the abundances measured in these
galaxies are very low, as has been observed in other low X-ray luminosity
ellipticals. Galaxy wind models greatly overestimate the abundance of the
gas. We look forward to further developments in galaxy modelling codes
which may be able to rectify these discrepancies.

\vspace{1cm}
\noindent{\textbf{Acknowledgments}\\
  The Authors would like to thank J.~C. Kempner, B.~J. Maughan and A.~M.
  Read for the use of their software and their advice on XMM and Chandra
  analysis, and D. Forbes for his help in the early stages of the project.
  We would also like to thank an anonymous referee whose comments
  significantly improved the paper. This work made use of the Digitized
  Sky Survey, the NASA/IPAC Extragalactic Database, and Starlink facilities
  at the University of Birmingham.  This research was supported in part by
  NASA grants NASA NAG5-10071 and NASA GO2-3186X, for which we are most
  grateful.

\bibliographystyle{mn2e}
\bibliography{../paper}

\begin{thebibliography}{}

\bibitem[\protect\citeauthoryear{{Acreman}, {Stevens}, {Ponman} \&
  {Sakelliou}}{{Acreman} et~al.}{2003}]{Acremanetal03}
{Acreman} D.~M.,  {Stevens} I.~R.,  {Ponman} T.~J.,    {Sakelliou} I.,  2003,
  MNRAS, 341, 1333

\bibitem[\protect\citeauthoryear{{Arnaud}, {Majerowicz}, {Lumb}, {Neumann},
  {Aghanim}, {Blanchard}, {Boer}, {Burke}, {Collins}, {Giard}, {Nevalainen},
  {Nichol}, {Romer} \& {Sadat}}{{Arnaud} et~al.}{2002}]{Arnaudetal02}
{Arnaud} M.,  {Majerowicz} S.,  {Lumb} D.,  {Neumann} D.~M.,  {Aghanim} N.,
  {Blanchard} A.,  {Boer} M.,  {Burke} D.~J.,  {Collins} C.~A.,  {Giard} M.,
  {Nevalainen} J.,  {Nichol} R.~C.,  {Romer} A.~K.,    {Sadat} R.,  2002, A\&A,
  390, 27

\bibitem[\protect\citeauthoryear{{Bacon}, {Monnet} \& {Simien}}{{Bacon}
  et~al.}{1985}]{Baconetal85}
{Bacon} R.,  {Monnet} G.,    {Simien} F.,  1985, A\&A, 152, 315

\bibitem[\protect\citeauthoryear{{Barnes}}{{Barnes}}{1989}]{Barnes89}
{Barnes} J.~E.,  1989, Nature, 338, 123

\bibitem[\protect\citeauthoryear{Bauer, Brandt, Sambruna, Chartas, Garmire,
  Kaspi \& Netzer}{Bauer et~al.}{2001}]{Baueretal01}
Bauer F.~E.,  Brandt W.~N.,  Sambruna R.~M.,  Chartas G.,  Garmire G.~P.,
  Kaspi S.,    Netzer H.,  2001, AJ, 122, 182

\bibitem[\protect\citeauthoryear{{Bender}}{{Bender}}{1988}]{Bender88}
{Bender} R.,  1988, A\&A, 202, L5

\bibitem[\protect\citeauthoryear{Bender, Surma, D\"{o}bereiner, Mollenhoff \&
  Madejsky}{Bender et~al.}{1989}]{Benderetal89}
Bender R.,  Surma P.,  D\"{o}bereiner S.,  Mollenhoff C.,    Madejsky R.,
  1989, A\&A, 217, 35

\bibitem[\protect\citeauthoryear{Blanton, Sarazin \& Irwin}{Blanton
  et~al.}{2001}]{Blantonetal01}
Blanton E.~L.,  Sarazin C.~L.,    Irwin J.~A.,  2001, ApJ, 552, 106

\bibitem[\protect\citeauthoryear{Brown \& Bregman}{Brown \&
  Bregman}{2001}]{Brownbreg01}
Brown B.~A.,  Bregman J.~N.,  2001, ApJ, 547, 154

\bibitem[\protect\citeauthoryear{{Buote}}{{Buote}}{2000a}]{Buote00}
{Buote} D.~A.,  2000a, ApJ, 539, 172

\bibitem[\protect\citeauthoryear{{Buote}}{{Buote}}{2000b}]{Buote00b}
{Buote} D.~A.,  2000b, MNRAS, 311, 176

\bibitem[\protect\citeauthoryear{{Buote}, {Lewis}, {Brighenti} \&
  Mathews}{{Buote} et~al.}{2003}]{Buoteetal03b}
{Buote} D.~A.,  {Lewis} A.~D.,  {Brighenti} F.,    Mathews W.~G.,  2003, ApJ,
  in press

\bibitem[\protect\citeauthoryear{Carollo, Franx, Illingworth \& Forbes}{Carollo
  et~al.}{1997}]{Carolloetal97}
Carollo C.,  Franx M.,  Illingworth G.,    Forbes D.,  1997, ApJ, 481, 710

\bibitem[\protect\citeauthoryear{Cash}{Cash}{1979}]{Cash79}
Cash W.,  1979, ApJ, 228, 939

\bibitem[\protect\citeauthoryear{Ciotti, D'Ercole, Pelegrini \& Renzini}{Ciotti
  et~al.}{1991}]{Ciottietal91}
Ciotti L.,  D'Ercole A.,  Pelegrini S.,    Renzini A.,  1991, ApJ, 376, 380

\bibitem[\protect\citeauthoryear{David, Forman \& Jones}{David
  et~al.}{1991}]{DavidForJo91}
David L.~P.,  Forman W.,    Jones C.,  1991, ApJ, 369, 121

\bibitem[\protect\citeauthoryear{Davis \& White}{Davis \&
  White}{1996}]{Daviswhite96}
Davis D.~S.,  White R. E.~I.,  1996, ApJ, 470, L35

\bibitem[\protect\citeauthoryear{Faber, Wegner, Burstein, Davies, Dressler,
  Lynden-Bell \& Terlevich}{Faber et~al.}{1989}]{Faberetal89}
Faber S.~M.,  Wegner G.,  Burstein D.,  Davies R.~L.,  Dressler A.,
  Lynden-Bell D.,    Terlevich R.~J.,  1989, ApJS, 69, 763

\bibitem[\protect\citeauthoryear{{Finoguenov} \& {Jones}}{{Finoguenov} \&
  {Jones}}{2000}]{FinoguenovJones00}
{Finoguenov} A.,  {Jones} C.,  2000, ApJ, 539, 603

\bibitem[\protect\citeauthoryear{{Forbes}, {Franx}, {Illingworth} \&
  {Carollo}}{{Forbes} et~al.}{1996}]{Forbesetal96}
{Forbes} D.~A.,  {Franx} M.,  {Illingworth} G.~D.,    {Carollo} C.~M.,  1996,
  ApJ, 467, 126

\bibitem[\protect\citeauthoryear{Forman, Jones \& Tucker}{Forman
  et~al.}{1985}]{Formanjonestucker85}
Forman W.,  Jones C.,    Tucker W.,  1985, ApJ, 293, 102

\bibitem[\protect\citeauthoryear{Garcia}{Garcia}{1993}]{Garcia93}
Garcia A.~M.,  1993, A\&AS, 100, 47

\bibitem[\protect\citeauthoryear{Gunn \& Gott}{Gunn \& Gott}{1972}]{GunnGott72}
Gunn J.~E.,  Gott J. R.~I.,  1972, ApJ, 176, 1

\bibitem[\protect\citeauthoryear{Helsdon \& Ponman}{Helsdon \&
  Ponman}{2000}]{Helsdonponman00}
Helsdon S.~F.,  Ponman T.~J.,  2000, MNRAS, 315, 356

\bibitem[\protect\citeauthoryear{Hibbard \& van Gorkom}{Hibbard \& van
  Gorkom}{1996}]{HibvanG96}
Hibbard J.~E.,  van Gorkom J.~H.,  1996, AJ, 111, 655

\bibitem[\protect\citeauthoryear{{Irwin}, {Athey} \& {Bregman}}{{Irwin}
  et~al.}{2003}]{IrAtheyBreg03}
{Irwin} J.~A.,  {Athey} A.~E.,    {Bregman} J.~N.,  2003, ApJ, accepted,
  astroph/0212422

\bibitem[\protect\citeauthoryear{Irwin, Sarazin \& Bregman}{Irwin
  et~al.}{2002}]{IrwinSarBreg02}
Irwin J.~A.,  Sarazin C.~L.,    Bregman J.~N.,  2002, ApJ, 570, 152

\bibitem[\protect\citeauthoryear{{Jansen}, {Lumb}, {Altieri}, {Clavel}, {Ehle},
  {Erd}, {Gabriel}, {Guainazzi}, {Gondoin}, {Much}, {Munoz}, {Santos},
  {Schartel}, {Texier} \& {Vacanti}}{{Jansen} et~al.}{2001}]{Jansenetal01}
{Jansen} F.,  {Lumb} D.,  {Altieri} B.,  {Clavel} J.,  {Ehle} M.,  {Erd} C.,
  {Gabriel} C.,  {Guainazzi} M.,  {Gondoin} P.,  {Much} R.,  {Munoz} R.,
  {Santos} M.,  {Schartel} N.,  {Texier} D.,    {Vacanti} G.,  2001, A\&A, 365,
  L1

\bibitem[\protect\citeauthoryear{{Kahabka}}{{Kahabka}}{2002}]{Kahabka02}
{Kahabka} P.,  2002, in {Lewin} W. H.~G.,  {van der Klis} M.,  eds, , Compact
  Stellar X-ray Sources.
Cambridge Universit Press

\bibitem[\protect\citeauthoryear{{Kraft}, {Nolan}, {Ponman}, C. \&
  {Raychaudhury}}{{Kraft} et~al.}{2003}]{Kraftetal03}
{Kraft} R.~P.,  {Nolan} L.~A.,  {Ponman} T.~J.,  C. J.,    {Raychaudhury} S.,
  2003, preprint

\bibitem[\protect\citeauthoryear{{Kundu}}{{Kundu}}{1999}]{Kundu99}
{Kundu} A.,  1999, Ph.D.~Thesis

\bibitem[\protect\citeauthoryear{{Lumb}}{{Lumb}}{2002}]{Lumb02}
{Lumb} D.,  2002, EPIC background files,
  http://xmm.vilspa.esa.es/docs/documents/CAL-TN-0016-2-0.ps.gz

\bibitem[\protect\citeauthoryear{Mackie \& Fabbiano}{Mackie \&
  Fabbiano}{1997}]{MackieFabb96}
Mackie G.,  Fabbiano G.,  1997, in Arnaboldi M.,  Da~Costa G.~S.,   Saha P.,
  eds, ASP Conf. Ser. 116: The Nature of Elliptical Galaxies; 2nd Stromlo
  Symposium Environmental and internal optical properties and the x-ray content
  of e and sos.
p.~401

\bibitem[\protect\citeauthoryear{{Malin} \& {Carter}}{{Malin} \&
  {Carter}}{1983}]{MalinCarter83}
{Malin} D.~F.,  {Carter} D.,  1983, ApJ, 274, 534

\bibitem[\protect\citeauthoryear{{Markevitch}}{{Markevitch}}{2002}]{Markevitch%
02}
{Markevitch} M.,  2002, preprint, astro-ph/0205333

\bibitem[\protect\citeauthoryear{{Marty}, {Kneib}, {Sadat}, {Ebeling} \&
  {Smail}}{{Marty} et~al.}{2002}]{Martyetal02}
{Marty} P.~B.,  {Kneib} J.~P.,  {Sadat} R.,  {Ebeling} H.,    {Smail} I.,
  2002, proc.~SPIE, 4851, 208

\bibitem[\protect\citeauthoryear{Matsumoto, Koyama, Awaki, Tsuru, Loewenstein
  \& Matsushita}{Matsumoto et~al.}{1997}]{Matsumotoetal97}
Matsumoto H.,  Koyama K.,  Awaki H.,  Tsuru T.,  Loewenstein M.,    Matsushita
  K.,  1997, ApJ, 482, 133

\bibitem[\protect\citeauthoryear{{Matsushita}, {Finoguenov} \& {B{\"
  o}hringer}}{{Matsushita} et~al.}{2003}]{Matsushitaetal03}
{Matsushita} K.,  {Finoguenov} A.,    {B{\" o}hringer} H.,  2003, A\&A, 401,
  443

\bibitem[\protect\citeauthoryear{Matsushita, Ohashi \& Makishima}{Matsushita
  et~al.}{2000}]{Matsushita00}
Matsushita K.,  Ohashi T.,    Makishima K.,  2000, PASJ, 52, 685

\bibitem[\protect\citeauthoryear{{Michard}}{{Michard}}{1998}]{Michard98}
{Michard} R.,  1998, A\&A, 335, 49

\bibitem[\protect\citeauthoryear{{Nieto}, {Bender} \& {Surma}}{{Nieto}
  et~al.}{1991}]{Nietoetal91}
{Nieto} J.,  {Bender} R.,    {Surma} P.,  1991, A\&A, 244, L37

\bibitem[\protect\citeauthoryear{Nulsen}{Nulsen}{1982}]{Nulsen82}
Nulsen P. E.~J.,  1982, MNRAS, 198, 1007

\bibitem[\protect\citeauthoryear{{O'Sullivan}, {Forbes} \&
  {Ponman}}{{O'Sullivan} et~al.}{2001a}]{OFP01cat}
{O'Sullivan} E.,  {Forbes} D.~A.,    {Ponman} T.~J.,  2001a, MNRAS, 328, 461

\bibitem[\protect\citeauthoryear{{O'Sullivan}, {Forbes} \&
  {Ponman}}{{O'Sullivan} et~al.}{2001b}]{OFP01age}
{O'Sullivan} E.,  {Forbes} D.~A.,    {Ponman} T.~J.,  2001b, MNRAS, 324, 420

\bibitem[\protect\citeauthoryear{{O'Sullivan}, {Ponman} \&
  {Collins}}{{O'Sullivan} et~al.}{2003}]{OPC03}
{O'Sullivan} E.,  {Ponman} T.~J.,    {Collins} R.~S.,  2003, MNRAS, 340, 1375

\bibitem[\protect\citeauthoryear{Pellegrini}{Pellegrini}{1994}]{Pellegrini94}
Pellegrini S.,  1994, A\&A, 292, 395

\bibitem[\protect\citeauthoryear{{Pellegrini}}{{Pellegrini}}{1999}]{Pellegrini%
99b}
{Pellegrini} S.,  1999, A\&A, 343, 23

\bibitem[\protect\citeauthoryear{Pellegrini \& Ciotti}{Pellegrini \&
  Ciotti}{1998}]{Pellegriniciotti98}
Pellegrini S.,  Ciotti L.,  1998, A\&A, 333, 433

\bibitem[\protect\citeauthoryear{{Pellegrini} \& {Fabbiano}}{{Pellegrini} \&
  {Fabbiano}}{1994}]{PellegriniFabbiano94}
{Pellegrini} S.,  {Fabbiano} G.,  1994, ApJ, 429, 105

\bibitem[\protect\citeauthoryear{{Pratt}, {Arnaud} \& {Aghanim}}{{Pratt}
  et~al.}{2001}]{Prattetal01}
{Pratt} G.~W.,  {Arnaud} M.,    {Aghanim} N.,  2001, in {Neumann} D.~M.,
  {Tranh Thanh Van} J.,  eds, Clusters of Galaxies and the High Redshift
  Universe Observed in X-rays: {XMM-Newton observations of galaxy clusters; the
  radial temperature profile of A2163}.
p. in press

\bibitem[\protect\citeauthoryear{Prugniel \& Simien}{Prugniel \&
  Simien}{1996}]{PrugnielSimien96}
Prugniel P.,  Simien F.,  1996, A\&A, 309, 749

\bibitem[\protect\citeauthoryear{Read \& Ponman}{Read \&
  Ponman}{1998}]{Readponman98}
Read A.~M.,  Ponman T.~J.,  1998, MNRAS, 297, 143

\bibitem[\protect\citeauthoryear{{Read} \& {Ponman}}{{Read} \&
  {Ponman}}{2003}]{ReadPonman03}
{Read} A.~M.,  {Ponman} T.~J.,  2003, A\&A, submitted

\bibitem[\protect\citeauthoryear{{Romanowsky}, {Douglas}, {Arnaboldi},
  {Kuijken}, {Merrifield}, {Napolitano}, M. \& {Freeman}}{{Romanowsky}
  et~al.}{2003}]{Romanowskyetal03}
{Romanowsky} A.~J.,  {Douglas} N.~G.,  {Arnaboldi} M.,  {Kuijken} K.,
  {Merrifield} M.~R.,  {Napolitano} N.~R.,  M. C.,    {Freeman} K.~C.,  2003,
  Science, 301, 1696

\bibitem[\protect\citeauthoryear{Sansom, Hibbard \& Schweizer}{Sansom
  et~al.}{2000}]{Sansom00}
Sansom A.~E.,  Hibbard J.,    Schweizer F.,  2000, AJ, 120, 1946

\bibitem[\protect\citeauthoryear{Sarazin, Irwin \& Bregman}{Sarazin
  et~al.}{2001}]{SarazinIrBreg01}
Sarazin C.~L.,  Irwin J.~A.,    Bregman J.~N.,  2001, ApJ, 556, 533

\bibitem[\protect\citeauthoryear{Schweizer \& Seitzer}{Schweizer \&
  Seitzer}{1992}]{Schweizerseitzers92}
Schweizer F.,  Seitzer P.,  1992, AJ, 104, 1039

\bibitem[\protect\citeauthoryear{{Scorza} \& {Bender}}{{Scorza} \&
  {Bender}}{1995}]{ScorzaBender95}
{Scorza} C.,  {Bender} R.,  1995, A\&A, 293, 20

\bibitem[\protect\citeauthoryear{Terlevich \& Forbes}{Terlevich \&
  Forbes}{2002}]{Terlevichforbes00}
Terlevich A.~I.,  Forbes D.~A.,  2002, MNRAS, 330, 547

\bibitem[\protect\citeauthoryear{{Tran}, {Tsvetanov}, {Ford}, {Davies},
  {Jaffe}, {van den Bosch} \& {Rest}}{{Tran} et~al.}{2001}]{Tranetal01}
{Tran} H.~D.,  {Tsvetanov} Z.,  {Ford} H.~C.,  {Davies} J.,  {Jaffe} W.,  {van
  den Bosch} F.~C.,    {Rest} A.,  2001, AJ, 121, 2928

\bibitem[\protect\citeauthoryear{{Trinchieri}, {Fabbiano} \&
  {Canizares}}{{Trinchieri} et~al.}{1986}]{Trinchierietal86}
{Trinchieri} G.,  {Fabbiano} G.,    {Canizares} C.~R.,  1986, ApJ, 310, 637

\bibitem[\protect\citeauthoryear{{van den Bergh}}{{van den
  Bergh}}{2001}]{vandenBergh01}
{van den Bergh} S.,  2001, PASP, 113, 154

\bibitem[\protect\citeauthoryear{{Weisskopf}, {Brinkman}, {Canizares},
  {Garmire}, {Murray} \& {Van Speybroeck}}{{Weisskopf}
  et~al.}{2002}]{Weisskopfetal02}
{Weisskopf} M.~C.,  {Brinkman} B.,  {Canizares} C.,  {Garmire} G.,  {Murray}
  S.,    {Van Speybroeck} L.~P.,  2002, PASP, 114, 1

\end{thebibliography}

\label{lastpage}

\end{document}